%% file: main.tex
\documentclass[5p, 9pt]{elsarticle}
\def\BibTeX{{\rm B\kern-.05em{\sc i\kern-.025em b}\kern-.08em
T\kern-.1667em\lower.7ex\hbox{E}\kern-.125emX}}

\usepackage{amsmath,amssymb,amsfonts}
\usepackage{textcomp}
\usepackage{xcolor}
\usepackage{booktabs}
\usepackage{siunitx}   
\usepackage{multirow}
\usepackage{multicol}
\usepackage{diagbox}
\usepackage[utf8]{inputenc}
\usepackage[T1]{fontenc}
\usepackage{hyperref}
\usepackage{url}         
\usepackage{nicefrac}
\usepackage{microtype}
\usepackage{algorithmic}

\usepackage{epsfig}
\usepackage{epstopdf}
\usepackage{graphicx}
\usepackage{latexsym}
\usepackage{comment}

\usepackage{mathrsfs}
\usepackage{float}
\usepackage{pifont}
\usepackage{yhmath}
\usepackage{amsthm}
\usepackage{wasysym}
\usepackage{array}
\usepackage[ruled]{algorithm2e}
\usepackage{bm}
\usepackage{changepage}
\usepackage{bigstrut,rotating}
\usepackage{caption} 
\usepackage{subcaption}
\usepackage{ulem}
\usepackage{colortbl}
\usepackage{lmodern}
\usepackage[T1]{fontenc}
\usepackage{titlesec}

\newtheorem{observation}{Observation}

\newcommand\figcaption{\def\@captype{figure}\caption}
\newcommand\tabcaption{\def\@captype{table}\caption}

\newcommand{\fakepar}[1]{\noindent\textit{#1}\qquad}

\usepackage{ifthen} 

\newcounter{todocounter}

\newcounter{todocounterA}
\renewcommand{\thetodocounterA}{\arabic{todocounterA}}
\newcommand{\todoA}[1]{%
  \refstepcounter{todocounterA}%
  \textcolor{blue}{[\textbf{TODO~\thetodocounterA}: \textrm{#1}]}%
}

\journal{Computer Networks}

\begin{document}
\begin{frontmatter}

\title{DRST: a Non-Intrusive Framework for Performance Analysis in Softwarized Networks}
%\todo{I think this paper is not ready for submission. Please hold on until further notice} 

%\title{Non-Intrusive and Automated Performance Intelligence via MLOps in Software Data Planes}

\author[1]{Qiong Liu}\ead{qiong.liu@ensea.fr}
\author[1]{Jianke Lin}\ead{jianke.lin@cyu.fr}
%\author[2]{Yuanyi Qiu}\ead{yuanyi.qiu@telecom-paris.fr}
\author[2]{Tianzhu~Zhang\corref{cor1}}\ead{tianzhu.zhang@nokia-bell-labs.com} %\cortext[cor1]{Corresponding author}
\author[3]{Leonardo~Linguaglossa}\ead{linguaglossa@telecom-paris.fr}

\address[1]{ETIS UMR 8051, CYU, CNRS, ENSEA, 95000 Cergy, France}
\address[2]{Nokia Bell Labs, 91300 Paris-Saclay, France}
\address[3]{LTCI, Telecom Paris, Institut Polytechnique de Paris, 91120 Palaiseau, France}

\begin{abstract}
The last decade has witnessed the proliferation of network function virtualization (NFV) in the telco industry, thanks to its unparalleled flexibility, scalability, and cost-effectiveness. However, as the NFV infrastructure is shared by virtual network functions (VNFs), sporadic resource contentions are inevitable. Such contention makes it extremely challenging to guarantee  the performance of the provisioned network services, especially in high-speed regimes (e.g., Gigabit Ethernet%\todoA{“We don’t do Terabit Ethernet.”}
). Existing solutions typically rely on direct traffic analysis (e.g., packet- or flow-level measurements) to detect performance degradation and identify bottlenecks; however, this approach is not always applicable due to significant integration overhead and system-level constraints.
%\todoA{This is not a problem of ‘per-packet’. This is more a ‘traffic analysis’ vs. ‘indirect readings’” Rather than relying on direct traffic analysis, we adopt indirect measurements based on low-level system signals.} 
This paper complements existing solutions with a lightweight, non-intrusive framework for online performance inference that easily adapts to \textit{drift} (i.e., a change over time of the actual state of our system). Instead of direct data-plane collection, we utilize hardware features in the underlying NFV infrastructure, introducing negligible interference in the data plane. Our Drift-Resilient and Self-Tuning (DRST) framework can be integrated into existing NFV systems with minimal engineering effort and operates without the need for predefined traffic models or VNF-specific customization. DRST is deployed via a lightweight MLOps pipeline that automates the adaptation under runtime drift. We show how DRST can deliver accurate performance inference or diagnose run-time bottleneck, as demonstrated through a comprehensive evaluation across diverse NFV scenarios.

%Through comprehensive evaluation across different high-speed NFV settings, our framework can effectively predict performance impairments using advanced Machine Learning (ML) models, identify performance bottlenecks via eXplainable AI (XAI) techniques, and maintain sustained accuracy under system evolution through an  MLOps-driven workflow that enables continuous monitoring \& retraining. \todoA{we leverage DRST maybe?”: through a DRST-compliant MLOps-driven workflow}\todo{Too vague, there are two claims to check: not sure about the "without requiring prior knowledge" (we still require a data collection or at least the labeling of some perf/topologies; also we don't do Terabit Ethernet (not even sure we can consider 10Gbps high speed anymore. We need to address this. The concept of drift and retraining should appear from the Abstract. }

\end{abstract}

\begin{keyword}
Network function virtualization, performance prediction, drift detection, non-intrusive, MLOps.
\end{keyword}

\end{frontmatter}

\input{paper/Sec1-Introduction}
\input{paper/Sec2-Background}

\input{paper/Sec3-Motivation}

\input{paper/Sec4-SystemDesignNew}

\input{paper/Sec5-Experiments}

\input{paper/Sec6-Conclusion}
\input{paper/appendix}

\bibliographystyle{elsarticle-num-names}
\bibliography{reference}
\end{document}

%% file: paper/Sec1-Introduction.tex
%!TEX root = ../main.tex

\section{Introduction}

%In recent years, the telco industry has undergone a profound transformation. With Network Function Virtualization (NFV) and Software-Defined Networking (SDN) at the forefront, network operators are pivoting away from traditional paradigms driven by proprietary, monolithic, and expensive middleboxes toward flexible, software-based equivalents running on commodity-off-the-shelf (COTS) servers. Meanwhile, high-speed I/O technologies, such as Intel
%Data Plane Development Kit (DPDK)~\cite{dpdk}, Mellanox Messaging Accelerator (VMA)~\cite{vma}, netmap~\cite{rizzo2012netmap}, extended Berkeley Packet Filter (eBPF)~\cite{ebpf}, Snabb~\cite{paolino2015snabbswitch}, and PF\_RING ZC~\cite{pf}, have greatly enhanced the capabilities of software packet processing, and contemporary software stacks manage to attain high performance regimes, e.g., 10/40/100~Gbps, which were previously exclusive to traditional hardware middleboxes~\cite{zhang2016opennetvm}. These breakthroughs in network softwarization are crucial for Communication Service Providers and cloud data centers, where performance, scalability, and cost-effectiveness are essential pillars of business value and revenue growth~\cite{zhang2020nfv}. %\footnote{\textbf{Leo note:} \emph{to check, as the problem is not "perpacket" or "perflow", but rather "via direct measurements / traffic measurements" versus "via indirect inference, through hardware counters". Actually it is orthogonal, we just don't look at the real traffic, rather at the effect of the traffic on the hardware.} Modified in the whole paper.}

In recent years, Network Function Virtualization (NFV) and Software-Defined Networking (SDN) have accelerated a shift in telco industry from proprietary, monolithic hardware appliances to agile, software-based functions deployable on commodity servers.  Meanwhile, high-speed I/O technologies, such as Intel
Data Plane Development Kit (DPDK)~\cite{dpdk}, Mellanox Messaging Accelerator (VMA)~\cite{vma}, netmap~\cite{rizzo2012netmap}, extended Berkeley Packet Filter (eBPF)~\cite{ebpf}, and Snabb~\cite{paolino2015snabbswitch}, have greatly enhanced the capabilities of software packet processing. Nowadays, modern software stacks can now attain high-performance regimes as 10/40/100~Gbps,  which were previously exclusive to traditional hardware middleboxes~\cite{zhang2016opennetvm}. These advances are vital for communication service providers and cloud data centers, where performance, scalability, and cost-efficiency underpin sustained growth and operational value~\cite{zhang2020nfv}.

Despite their popularity, softwarized networks also face some intrinsic obstacles in practice. In particular, compared to traditional hardware middleboxes that utilize dedicated circuits for packet processing, the software data plane is more susceptible to performance impairments due to the shared nature of the underlying virtual infrastructure~\cite{wu2015perfsight} and various bottlenecks~\cite{cai2021understanding}. The colocated VNFs can compete for subsystem resources, which severely degrades the quality of network services. Therefore, performance prediction and analysis constitute the first fundamental step to fulfilling SLAs~\cite{manousis2020contention}.
Existing solutions mainly combine direct feature collection with statistical reasoning~\cite{gong2020microscope,haecki2022diagnose, arashloo2023formal} or machine learning~\cite{yao2022aquarius, bronzino2021traffic,wan2024cato} for fine-granular, accurate performance analysis in high-speed networks. However, these solutions are not always applicable due to: (i) the potentially substantial instrumentation overhead, especially for VNFs from heterogeneous vendors~\cite{li2023lemonnfv}, (ii) the collateral interference on the high-speed data plane, which can damage both network performance and data fidelity~\cite{zhang2019flowatcher}, (iii) inflexible inference without trade-offs in predictivity, inference latency, and interpretability, which are crucial CSP requirements~\cite{liu2024operationalizing}, %\footnote{\textbf{Leo Note:} \emph{unclear, we need to explicit this point} clarify in section 5.2.}
(iv) the ensuing data drift and model decay as the network system evolves, making it strenuous for heuristic and data-driven methods to sustain accuracy~\cite{yang2021quality}.

This paper complements existing solutions with a novel framework for performance intelligence in high-speed softwarized networks. Instead of relying on direct traffic measurement for data collection, we leverage the low-level hardware features of different subsystems, such as the CPU pipeline, multi-level caches, Random Access Memory (RAM), and Input/Output (I/O). These features are ubiquitous in modern COTS servers and can be acquired with standard profiling tools~\cite{perf, pcm, vtune}. Although less relevant than the packet-/flow-level statistics, these features embody rich run-time network information and can be combined with data-driven analytics to deliver actionable insights~\cite{shelbourne2019learnability,shelbourne2021inference, liu2024non}.
%\todoA{To check.}
We employ advanced ML algorithms for performance prediction (e.g., step-ahead KPI forecasting) and analysis (e.g., bottleneck detection). The main contributions of this work are as follows: 
\begin{itemize}
\item \textit{Non-intrusive data collection}: Our framework leverages the low-level hardware features for analytics, which incurs  negligible overhead on the data plane.
\item \textit{Seamless integration}: Our framework can seamlessly integrate with existing NFV systems without prior domain knowledge and extra engineering overhead. 
%\item \textit{Accurate inference}: We conduct extensive feature analysis and ML model exploration for accurate performance prediction and analysis. Specifically, we implement two neural network architectures to capture the complex, non-linear data patterns and network dynamics.  \todo{At some point you talk about accurate inference, or "variable accuracy" or something like tradeoff with respect to other metrics. It should be clarified. } '\todoA{drift, achieving robust inference even under evolving traffic patterns and system drift. } 

\item \textit{Model benchmarking and robust selection}: We evaluate several ML architectures across heterogeneous scenarios and retain those that balance accuracy and latency. %Through analysis across diverse traffic regimes, we identify MLP and LSTM as robust choices balancing accuracy and runtime compatibility. 

%\item \textit{Model benchmarking and robust selection}: We evaluate multiple ML models across heterogeneous NFV scenarios involving varied VNFs, topologies, and stimulus patterns. We select architectures that balance accuracy, generalization, and inference latency.

%\footnote{\textbf{Leo Note:} \emph{in some point you talk about accurate inference, or "variable accuracy" or something like tradeoff with respect to other metrics. It should be clarified.} modified  }

\item \textit{End-to-end pipeline with DRST}: We implement our solution as an end-to-end ML pipeline with Drift Resilient and Self-Tuning (DRST) capabilities, enabling continuous monitoring and adaptive retraining to sustain accuracy under data drift.
\end{itemize}

This  paper is organized as
follows: We first review relevant background and prior work in Section~\ref{sec:background}, then outline our motivations in Section~\ref{sec:motivation}. The proposed system design is described in Section~\ref{sec:design}, and experimental performance is evaluated in Section~\ref{sec:exp}. Section~\ref{sec:conclusion} concludes the paper. %Section~\ref{sec:background} presents the background and related work. Section~\ref{sec:motivation} outlines key motivations. The system design is described in Section~\ref{sec:design}, followed by experimental results and analysis in Section~\ref{sec:exp}. Finally, we conclude the paper in Section~\ref{sec:conclusion}. 

%% file: paper/Sec2-Background.tex
%!TEX root = ../main.tex

\section{Background}
\label{sec:background}

\subsection{High-speed softwarized networks}

Traditionally, software-based networking solutions, such as the Click Modular Router~\cite{kohler2000click}, have been primarily used for fast prototyping and functional testing due to their unparalleled accessibility, flexibility, and customizability. However, specialized hardware equipment (or middleboxes) stood out in real-world deployments thanks to their far superior packet processing capabilities. In recent years, the rapid development of software acceleration techniques, such as kernel-bypass, poll-mode, batch processing, and parallel computing, has significantly narrowed the "performance gap" between software solutions running on COTS servers and specialized middleboxes~\cite{zhang2020nfv}. Nowadays, software packet processing is an integral part of the modern telco industry~\cite{dpdk}.

However, softwarized networks still bear several inherent limitations. Co-located Virtual Network Functions (VNFs) suffer performance impairments due to the erratic contentions in the shared NFV infrastructure~\cite{manousis2020contention}. Consequently, network operators are forced to spend a lot of time pinpointing and resolving performance issues~\cite{gong2020microscope}. Such problems are intricate to predict due to the voluminous and heterogeneous traffic therein. The growing complexity of the NFV systems and network services further compounds the situation~\cite{zheng2020nfv}. As modern COTS servers continue to gain new functionalities, the software data plane can encompass numerous configuration knobs, including hardware options and software parameters. This vast search space makes it extremely challenging to anticipate and prevent performance contentions.
In parallel, network service structures are transforming beyond the conventional linear service function chains (SFCs), and many research efforts strive for enhanced service provisioning by parallelizing VNFs as Directed Acyclic Graphs (DAGs)~\cite {li2023lemonnfv,sun2017nfp, zhang2017parabox, lin2018dag}. As detailed in~\cite{sun2017nfp}, $53.8\%$ of VNF pairs in enterprise networks were parallelizable. Such flexible compositions increase the difficulty of diagnosing performance issues across service and infrastructure layers~\cite{gong2020microscope}.
In essence, there is an urgent need for novel approaches to accurately predicting performance degradations and identifying the root causes of complex network services. %\todoA{unclear.}

%Such convoluted interactions across multiple infrastructure and service layers make it exceedingly challenging to foresee the occurrence of performance issues~\cite{gong2020microscope}.

\subsection{Resource contention in software data plane}
\begin{figure}[t]
\centering
\includegraphics[width=.47\textwidth]{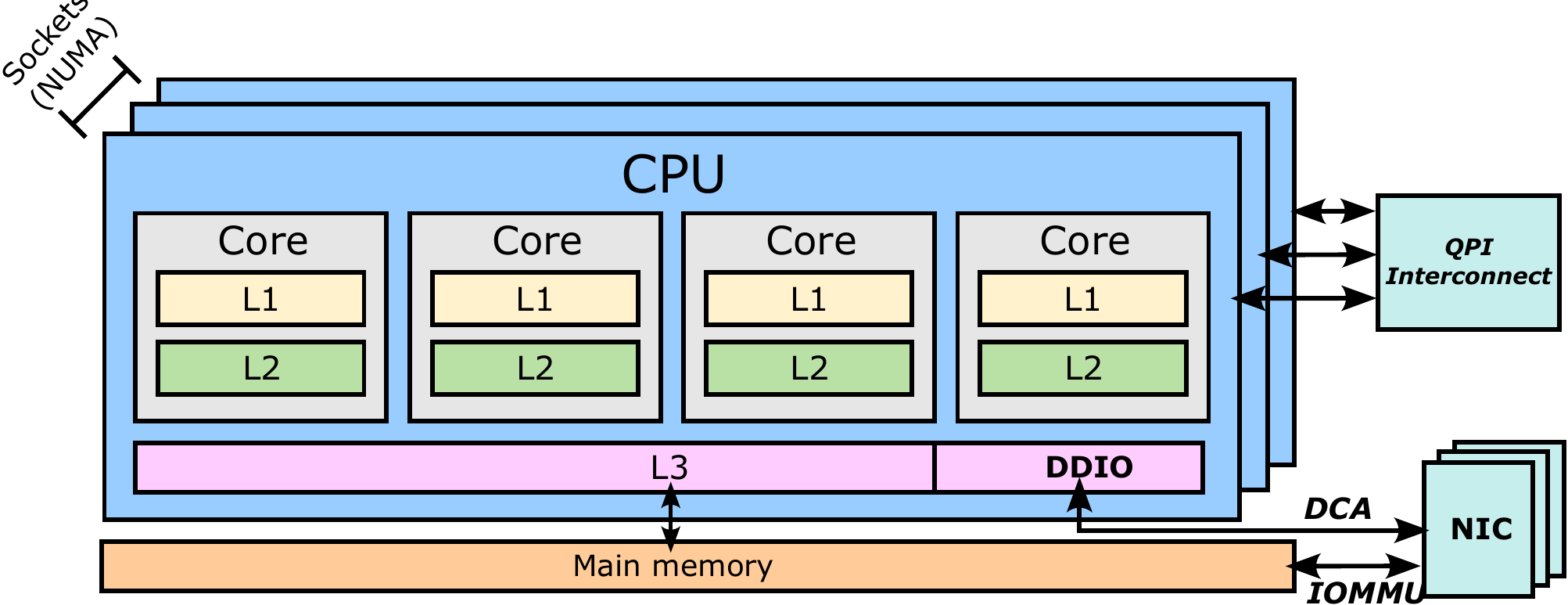}
\caption{The architecture of a modern COTS server.}
\label{fig:cots}
\end{figure} 
%\footnote{\todo{I think that the figure 1 is not 100\% accurate, I attach a new corrected version that reflects more the reality. We need to show the NUMA nodes (different sockets) as well as the QPI to interconnect them via the I/O controller. If you agree, you can just replace it. }}
%Modern COTS servers adopted in NFV deployments are typically multi-socket NUMA systems equipped with multi-level cache hierarchies and support for kernel-bypass packet I/O. \textcolor{orange}{As illustrated in Fig.~\ref{fig:cots}, each socket connects to its local memory and I/O bus, and contains multiple CPU cores with private L1 and L2 caches. Cores on the same NUMA node share a larger Last-Level Cache (LLC), while inter-node communication occurs through interconnects such as Intel\textsuperscript{\textregistered} QuickPath Interconnect (QPI).}\todoA{adapt to new figure.}

%Modern COTs servers are commonly multi-processor systems with multiple Non-Uniform Memory Access (NUMA) sockets, each with local memory and I/O buses to alleviate memory access contention. To further enhance data/instruction locality, COTS servers employ a multi-level cache design.  As shown in Fig.~\ref{fig:cots}, each CPU is equipped with multiple cores, with each having its own built-in L1 caches (for both data and instruction) and L2 caches. Cores located within the same NUMA node share the same L3 cache with a much larger size than L1/L2 caches. 

Modern COTS servers rely on multi-socket NUMA architectures, where each socket has access to local memory and I/O buses, thereby reducing latency from remote accesses. To support efficient memory usage and reduce bottlenecks, such systems adopt a hierarchical cache design. Fig.~\ref{fig:cots} shows that each CPU contains multiple cores, with its own built-in L1 and L2 caches. Cores located within the same NUMA node share the same L3 cache, which has a much larger size than the L1/L2 caches. 
Incoming packets at the Network Interface Controller (NIC) may be routed directly to the main memory through the Input-Output Memory Management Unit (IOMMU) or to the L3 cache via Direct Cache Access, e.g., the Intel\textsuperscript{\textregistered}  Direct Data I/O (DDIO). Cores on different NUMA nodes can only communicate via a specialized interconnect, such as the Intel\textsuperscript{\textregistered} QuickPath Interconnect (QPI).  

Given the complex layout of various components and the intricate interactions of the COTS servers, software packet I/O operations can still suffer from various contentions:

\begin{itemize}
\item \textit{CPU share}: The allocated CPU share directly decides how fast packets can be processed. Despite the high frequency of modern CPUs, performance variance can still emerge due to dynamic frequency scaling (e.g., Intel\textsuperscript{\textregistered} Turbo boost). Additionally, non-NFV workloads can be allocated to a VNF's cores due to misconfigured scheduling policies, resulting in performance losses. CPU isolation mechanisms, e.g., Linux \texttt{isolcpus}, can only alleviate the issue. 
\item  \textit{Multi-level caches}:  %While cache accesses are much faster than main memory, multi-level caches have been widely identified as a major performance bottleneck~\cite{dobrescu2012toward}. NFV frameworks often distribute packet processing across multiple cores, leading to severe contention for the Last-Level Cache (LLC). Cache partitioning techniques, such as Intel\textsuperscript{\textregistered} Cache Allocation Technology, cannot always prevent this. For instance, large incoming packets may contend for  Direct Data I/O (DDIO), causing the so-called leaky Direct Memory Access (DMA) problem~\cite{tootoonchian2018resq}. Given the architectural diversity across CPU generations, jointly optimizing LLC and DDIO remains a challenging task, especially in cloud environments~\cite{yuan2021don}.
While cache accesses are way faster than main memory, prior works have widely deemed multi-level caches the major performance bottleneck~\cite{dobrescu2012toward}. Many NFV frameworks streamline packet processing across multiple cores, which can cause severe contention for the Last-level caches (LLCs). Cache partitioning techniques, such as Intel\textsuperscript{\textregistered} Cache Allocation Technology (CAT), cannot always prevent such contentions, e.g., large incoming packets can contend for DDIO, which is referred to as the leaky Direct Memory Access (DMA) problem~\cite{tootoonchian2018resq}. As the caching systems of modern CPU microarchitectures differ across generations, jointly optimizing LLC and DDIO remains a daunting task, especially in cloud data centers~\cite{yuan2021don}.

\item \textit{Memory bandwidth}: Contention for memory bandwidth further slows the packet path. For instance, VNFs with lower LLC shares can incur high cache misses, which saturate the memory bandwidth for all the co-located VNFs and network services~\cite{chintapalli2023nfvpermit}. 
\end{itemize}

Other bottlenecks also exist. For instance, packet I/O across multiple NUMA nodes can be extremely slow due to the QPI contention~\cite{niu2017unveiling}, which can be avoided with a NUMA-aware design.

%\todo{Background: globally ok, but lots of acronyms are used once or never used. There are duplicates, I highlighted what I could find, but there may be more. }

\subsection{Related work}\label{sec:related}
Existing solutions commonly employ direct, per-packet measurement for data and feature collection, as well as performance analysis. For instance,  NFVPerf~\cite{naik2016nfvperf} applied packet mirroring for feature collection. PPTMon~\cite{van2021pptmon} employed event filtering and timestamp embedding to monitor the processing latency of VNFs. However, these works involve expensive operations in the software data plane and cannot cater to high-speed networks that handle millions of packets per second. 
Such tools fail to handle the huge input load in high-speed networks. For example, the throughput can reach $14.88$ Mpps with the end-to-end network service latency in sub-microseconds for $64$-byte synthetic packets on a $10$ Gbps link.
Some solutions were designed for high-speed regimes but were subject to enormous integration overhead, huge resource footprint, or operational constraints. For instance, NFV-VIPP~\cite{dodare2019nfv} captured the execution states of DPDK-augmented VNFs but required manually attaching the per-VNF threads, a nontrivial operational exertion in production networks. Additionally, each monitoring thread must occupy one CPU core, which was costly given the limited number of cores on COTS servers~\cite{zhang2019benchmarking,zhang2019comparing,zhang2021performance}.
Microscope~\cite{gong2020microscope} deduced performance bottlenecks via the runtime queuing states, which were not always obtainable in real networks~\cite{lan2016embark}. Moreover, as VNFs can originate from heterogeneous vendors, code instrumentation and customization are necessary, which can be highly burdensome due to the diversified implementation paradigms and operational patterns~\cite{li2023lemonnfv}.

To circumvent these obstacles, a line of work explored the low-level features and data-driven algorithms to derive performance insights. In particular, Dobrescu et al.~\cite{dobrescu2012toward} extrapolated the performance of software data planes using the cache features. Shelbourne et al.~\cite{shelbourne2021inference,shelbourne2019learnability} inferred throughput and packet losses for DPDK-based VNFs by exploring a larger set of low-level features. Still, they only analyzed the impact of input traffic without considering other system-level performance interferences.
Antonis et al.~\cite{manousis2020contention} developed SLOMO, a multivariable performance prediction framework to investigate common system-level contentions and employ gradient-boosting regression to predict service throughput.
Although these works delivered promising outcomes, they were primarily designed for singleton VNFs, without considering end-to-end network services (e.g., SFCs) that comprise multiple VNFs with varying topological compositions. Furthermore, these solutions were only tested in controlled environments. They did not consider the practical challenges of deploying AI/ML in real network systems with much higher scale, complexity, and dynamism~\cite{liu2024operationalizing}, which makes it hard to achieve long-term, sustainable accuracy. In particular, given the data-driven nature of ML-based solutions, fulfilling performance guarantees in the presence of data and environmental drifts is non-trivial~\cite{yang2021quality}. For the sake of clarity, we summarize in Table~\ref{table:acronyms} the main acronyms used in the remainder of this paper.

{\begin{table}[!tb]
\footnotesize
\centering
\begin{tabular}{l||l}
\toprule
\textbf{Acronym} & \textbf{Definition} \\
\midrule
COTS  & Commodity Off-The-Shelf \\
DAG   & Directed Acyclic Graph \\
DDIO & Direct Data I/O  \\
DirREC & Direct Recursive Forecasting Strategy \\
DMA & Direct Memory Access \\
DRST& Drift-Resilient and Self-Tuning \\
DPDK  & Data Plane Development Kit \\
eBPF&extended Berkeley Packet Filter  \\
%GBR& Gradient Boosting Regression \\
KPI   & Key Performance Indicator \\
IOMMU & Input–Output Memory Management Unit\\
LSTM  & Long Short-Term Memory \\
LLC& Last-Level Cache \\
MANO  & Management and Network Orchestration \\
MAPE& Mean Absolute Percentage Error\\
%& Mellanox Messaging Accelerator \\
MLP& multi-layer perceptron\\
NIC & Network Interface Controller \\
NFV   & Network Function Virtualization \\
PTP   & Precision Time Protocol \\
PCM   & Performance Counter Monitor \\
PMU & Performance Monitoring Units \\
%QPI & QuickPath Interconnect \\
RAM   & Random Access Memory \\
%VMA   & Mellanox Messaging Accelerator \\
VNF   & Virtual Network Function \\
SDN& Software-Defined Networking \\
SFC   & Service Function Chain \\
SHAP  & SHapley Additive exPlanations \\
\bottomrule
\end{tabular}
\caption{List of acronyms}
\label{table:acronyms}
\end{table}
}

%% file: paper/Sec3-Motivation.tex
\section{Motivation}
\label{sec:motivation}

This section presents an observational study on non-intrusive data collection. We first compare the overhead of direct versus indirect measurements, and then analyze how low-level features relate to KPIs under varied traffic and service settings.

\begin{figure}[!tb]
\small
    \centering
    \begin{subfigure}{0.23\textwidth}
        \centering
        \includegraphics[width=\textwidth]{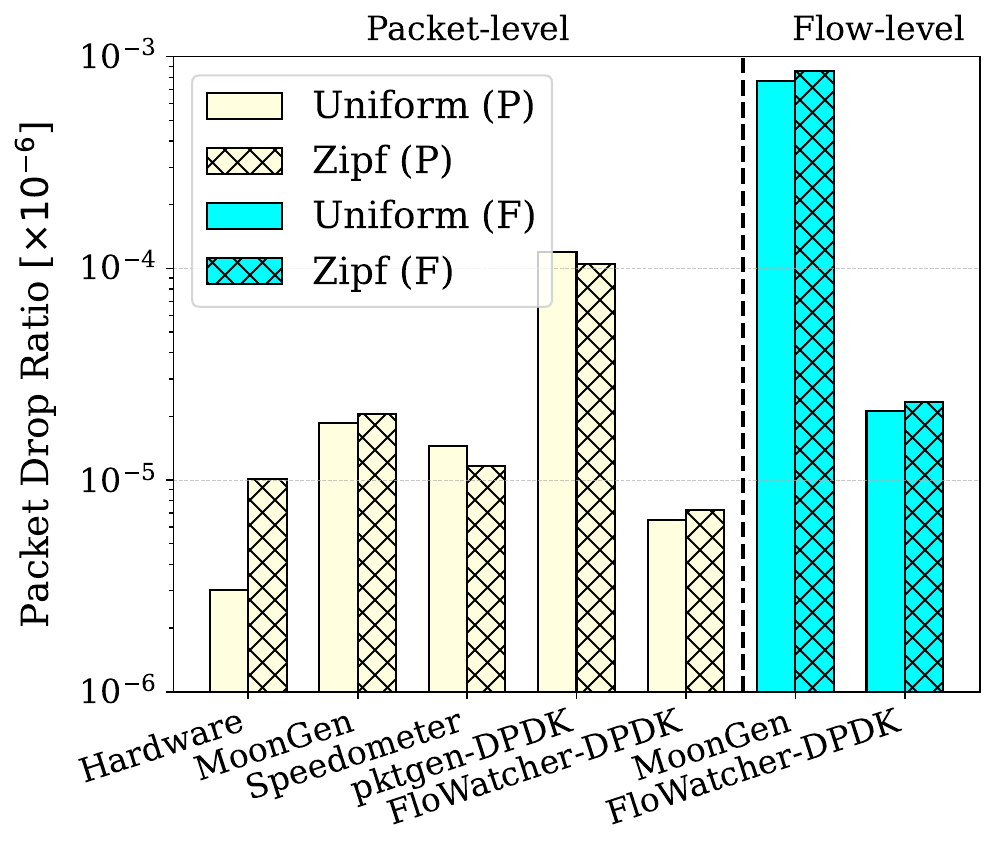}
        \caption{Open-loop overhead}
        \label{fig:open-loop-overhead}
    \end{subfigure}
    \begin{subfigure}{0.24\textwidth}
        \centering
        \includegraphics[width=\textwidth]{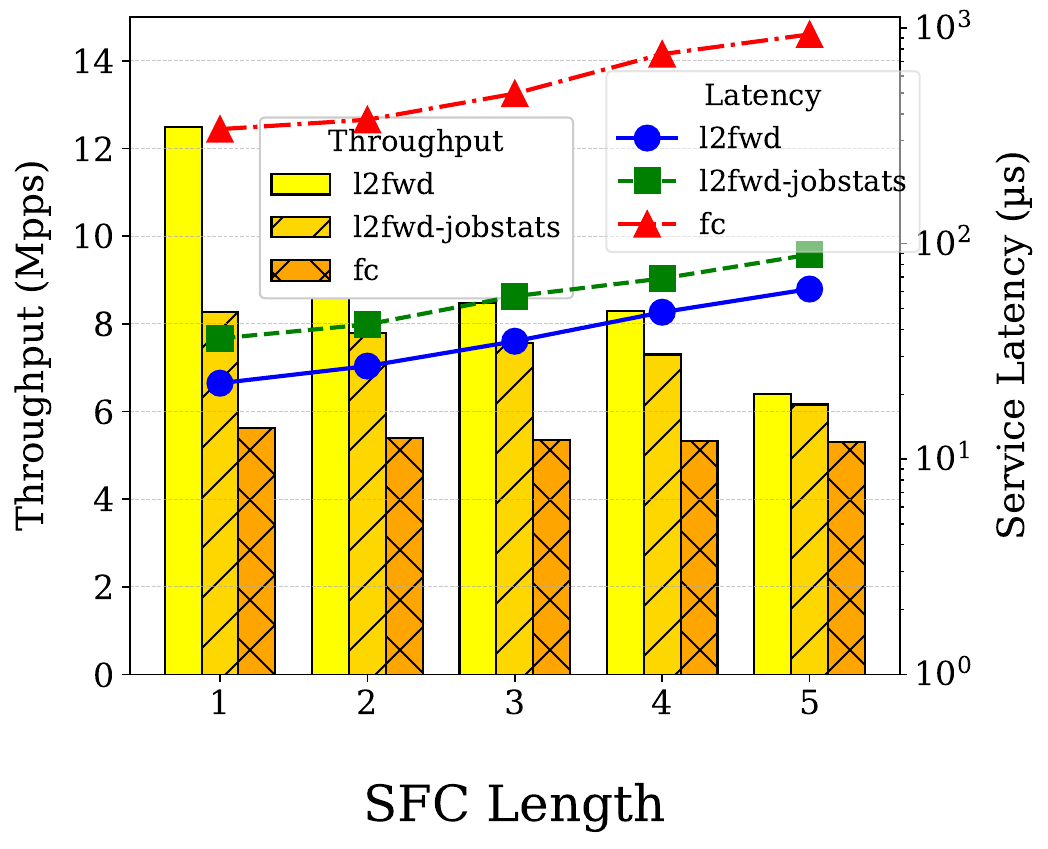}
        \caption{Closed-loop overhead}
        \label{fig:closed-loop-overhead}
    \end{subfigure}
    \caption{Overhead of direct  vs. indirect  measurements}
    \label{fig:overhead}
\end{figure}

%\subsection{Per-packet data collection overhead}
 \subsection{Direct vs. Indirect overhead}
To evaluate the runtime overhead introduced by different monitoring approaches, we compare direct measurements (e.g., packet- and flow-level analysis) with indirect measurements via hardware performance counters. To illustrate the overhead of per-packet data collection, we consider the de facto test scenarios for network measurement: open-loop and closed-loop \cite{zhang2018flowmon}. 
For the open-loop test, we generate 64B of traffic at 10 Gbps and deploy four state-of-the-art network measurement tools, i.e., MoonGen~\cite{emmerich2015moongen}, Speedometer~\cite{speedometer}, pktgen-DPDK~\cite{pktgen}, and FloWatcher-DPDK~\cite{zhang2019flowatcher}, to measure the throughput at the receiving end. MoonGen and FloWatcher-DPDK are also configured to collect per-flow statistics (e.g., flow size and inter-arrival gaps). The synthetic traffic consists of 60K flows, whose sizes are configured to follow Uniform and Zipf distributions. As illustrated in Fig.~\ref{fig:open-loop-overhead}, per-packet measurement incurs non-negligible overhead, even if we only access the NIC's counters (i.e., Hardware in the figure). When we proceed to perform packet- and flow-level measurements, the overhead only increases. Although the packet loss ratios seem small ($10^{-5}$-$10^{-4}$), this can already cause severe issues, given the high input rates of millions of packets per second in high-speed networks.

We deploy a sample SFC comprising identical VNFs for the closed-loop test, each of which performs Layer 2 packet forwarding. The VNFs are containerized with Docker and interconnected using FastClick~\cite{barbette2015fast}. We chose three VNFs from the DPDK example library, i.e., {\em l2fwd, l2fwd-jobstats,} and {\em flow classification (fc)}. l2fwd performs simple forwarding. l2fwd-jobstats and fc further collect packet- and flow-level statistics, respectively. Similar to the open-loop measurement, we continuously inject 64B of synthetic traffic to the SFC at the line rate and measure the throughput. We also send Precision Time Protocol (PTP) packets to measure the end-to-end latency. The SFC length is varied from 1 to 5~\footnote{Note that the performance deteriorates as the SFC gets longer, mainly due to the accumulated memory copying and inter-core communication overhead~\cite{zhang2019comparing}.}. As illustrated in Fig.~\ref{fig:closed-loop-overhead}, per-packet measurement causes significant performance degradation. In particular, fc causes the throughput to drop by up to $50\%$ while extending the latency by one order of magnitude. %\todoA{highlight do not know why}

To demonstrate the advantage of our approach, we repeat both experiments, but run \textit{perf} to collect low-level features every $100$\,ms. Hardware counters reflect averaged or accumulated metrics over each sampling interval. Note that shorter intervals may be necessary for detecting transient phenomena such as jitter or microbursts, they are less critical for our target of sustained performance inference in NFV environments. To test different levels of interactions, we sequentially assign perf to the same worker cores as the VNFs, to different idle cores on the same NUMA node, and to cores on the other NUMA node. In all cases, the perceived throughput and latency remain the same.  

%Since hardware counters reflect aggregated behaviors over the sampling window, the sampling interval must balance resolution and system overhead. Shorter intervals offer finer detail but may interfere with execution, while longer intervals reduce overhead but may smooth out transient effects. In our case, the selected intervals (e.g., 100\,ms) are sufficient to capture relevant performance trends without affecting system behavior.\footnote{\textbf{Leo Note} \emph{100ms as sampling period was used due to performance issues and also as it was just okay. However, they may question why 100ms and not another value. BTW, if we perform optimization or inference every 100ms, at 100Gbps we lose 10 Gb worth of traffic. I think we should detail this point, to say that we don't do anomaly detection or intrusion prevention, where every packet counts. } clarified.}
 %\todoA{Very important. At 200 Gbps, we "miss" 20 Gb worth of traffic!}

%\begin{observation}
%Data collection via PMUs imposes negligible overhead on the data path compared to most existing solutions based on per-packet data collection. 
%\end{observation}

\begin{observation}
Indirect measurement via PMUs provides a non-intrusive and low-overhead alternative to conventional direct measurement approaches.
\end{observation}

\subsection{Feature relevance analysis: KPIs}\label{sec:correlated_metrics}
\begin{figure}[!tb]
\begin{center}
\includegraphics[width=0.48\textwidth]{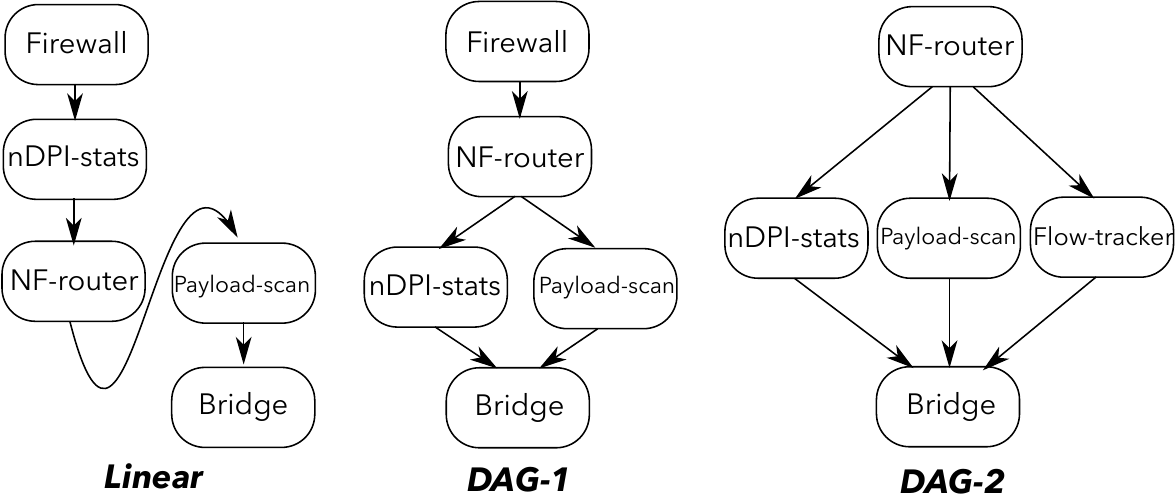}
\caption{Three typical network service topologies}
\label{fig:sfcs}
\end{center}
\end{figure}

\begin{table*}[!tb]
\scriptsize
\centering
\begin{subtable}{0.4\linewidth}
\centering
\begin{tabular}{l||c|c|c|c|c||c} 
\toprule
\diagbox{\textbf{Features}}{\textbf{VNF}} & Bridge& Payload-scan&NF-router& nDPI  &Firewall & Average \\
\midrule
LLC-load & 0.94 & 0.98 & 0.97 & 0.98 & 0.96 & 0.97 \\
Cache-reference & 0.95 & 0.97 & 0.97 & 0.98 & 0.97 & 0.94 \\
LLC-stores & 0.97 & 0.96 & 0.96 & 0.97 & 0.97 & 0.97 \\
L1-Dcache-load-misses & 0.95 & 0.97 & 0.97 & 0.98 & 0.97 & 0.97 \\
Instructions & 0.79 & 0.92 & 0.86 & 0.78 & 0.89 & 0.92 \\
Branches & 0.79 & 0.92 & 0.87 & 0.79 & 0.89 & 0.83 \\
Mem-stores & 0.39& 0.50& 0.42& 0.90&0.53& 0.55\\
\hline
Cache Misses & 0.32 & 0.18 & 0.35 & 0.65 & 0.36 & 0.38 \\
Cycles & 0.14 & 0.08 & 0.14 & 0.14 & 0.06 & 0.12 \\
\bottomrule
\end{tabular}
\caption{Throughput}
\label{table:correlation_load_stimulus} 
\end{subtable}
\hfill
\begin{subtable}{0.4\linewidth}
\centering
\begin{tabular}{l||c} 
\toprule
\multirow{2}{*}{\textbf{Features}} & \multirow{2}{*}{\textbf{Average}} \\
& \\
\midrule
LLC-load & 0.59 \\
Cache-reference & 0.58 \\
LLC-stores & 0.57 \\
L1-Dcache-load-misses & 0.55 \\
\hline
Instructions & 0.48 \\
Branches & 0.47 \\
Mem-stores & 0.11 \\
Cache Misses & 0.13 \\
Cycles & 0.02 \\
\bottomrule
\end{tabular}
\caption{Latency}
\label{table:latency_load_stimulus}
\end{subtable}
\caption{Correlated features with throughput and latency under load stimulus}
\label{tab:comparison-load-stimulus}
%\end{adjustwidth}
\end{table*}

\begin{table*}[!htb]
\centering
\begin{subtable}{0.4\linewidth}
\centering\scriptsize
\begin{tabular}{l||c|c|c|c|c||c} 
\toprule
\diagbox{\textbf{Features}}{\textbf{VNF}} & Bridge& Payload-scan&NF-router& nDPI  &Firewall & Average \\
\midrule
LLC-load&0.80&0.67&0.55&0.54&0.44&0.60\\
Cache-reference&0.81&0.73&0.58&0.46&0.45& 0.61\\
LLC-stores & 0.80&0.74&0.60&0.45&0.32 &0.58\\
L1-dcache-load-misses & 0.81&0.72&0.58&0.45&0.45&0.60\\
\hline
Cycles & 0.35&0.30&0.29&0.25&0.22&0.28\\
Instructions &0.21&0.21&0.24&0.21&0.19&0.21\\
Branches&0.21&0.21&0.24&0.21&0.19&0.21\\
Mem-stores & 0.12& 0.08& 0.04&0.73& 0.03& 0.20\\
Cache Misses &0.00&0.06&0.04&0.03&0.05&0.03\\
\bottomrule
\end{tabular}
\caption{Throughput}
\label{table:correlation_resource_stimulus} 
\end{subtable}
\hfill
\begin{subtable}{0.4\linewidth}
\centering\scriptsize
\begin{tabular}{l||c} 
\toprule
\multirow{2}{*}{\textbf{Features}} & \multirow{2}{*}{\textbf{Average}} \\
& \\
\midrule
LLC-load & 0.31 \\
Cache-reference & 0.23 \\
LLC-stores & 0.18 \\
Branches & 0.17 \\
L1-Dcache-load-misses & 0.15 \\
Cycles & 0.14 \\
Instructions & 0.13 \\
Mem-stores & 0.12\\
Cache Misses & 0.02 \\
\bottomrule
\end{tabular}
\caption{Latency}
\label{table:latency_resource_stimulus}
\end{subtable}
%\end{adjustwidth}
\caption{Correlated features with throughput and latency under resource stimulus}
\label{tab:comparison-resource-stimulus}
\end{table*}

\textit{Perf} exposes hundreds of hardware features, but many are noisy for ML. We must identify a refined subset of expressive features with strong predictive power.

To identify the relevant hardware features, we construct three typical network service topologies (or SFCs): a linear service chain and two DAGs, as shown in Fig.~\ref{fig:sfcs}. The VNFs ({\em Firewall}, {\em nDPI-stat}, {\em NF-router}, {\em Bridge,} {\em Payload-scan},  and {\em Flow-tracker}) are implemented and open-sourced by ONVM developers. We then inject traffic at random rates and inspect the tendencies of the features.  To represent typical Internet traffic, we configure MoonGen to generate IMIX traffic consisting of a variety of packet sizes in the ratio 64B: 570B: 1514B = 7: 4: 1. We collect the low-level features and performance metrics for different service topologies under both load and resource stimuli tests. 

We utilize Pearson's correlation and mutual information to assess the statistical dependencies between the collected features and the KPIs. We keep features whose correlation with a KPI exceeds 0.5. Tables~\ref {tab:comparison-load-stimulus} and~\ref {tab:comparison-resource-stimulus} list the correlated features with different KPIs for the linear SFC. The results are coherent with our tendency observations. There is no dominant feature that consistently achieves the highest correlation across different VNFs, suggesting the joint impact of multiple features on SFC performance. Note that similar feature correlations have also been observed with the DAG topologies; we omitted them for space sake.
%\begin{observation}
%Throughput correlation: When considering different network stimuli, distinct features are more suitable for capturing their characteristics. No VNF emerges as each feature's dominant indicator in loaded stimulus experiments. 

\fakepar{Correlation with throughput}
Under load stimulus, the features in the Table.~\ref{table:correlation_load_stimulus} shows consistently high correlations with the throughput across VNFs. %, which suggests that the input rate uniformly impacts the entire SFC. 
In particular, cache-related features, especially cache-reference rate and L1-dcache-load-misses, strongly correlate with the throughput. 
% A clear correlation is observed between the features of VNFs positioned at the end of the SFC and the throughput. Notably, this correlation gradually weakens as the distance between a VNF's position in the SFC and the output NIC.
In contrast, as shown in Table~\ref{table:correlation_resource_stimulus}, the correlation of those same features under resource stimulus exhibits an ascending pattern: the correlation is small at the beginning of the chain and increases towards the end. The bridge VNF (last column) shows the highest correlation because it is memory-intensive.
We observe that certain VNF-specific behaviors contribute to the peculiarities of individual features. For instance, nDPI's mem-stores feature, as shown under both stimulus tests, exhibits a very high correlation with throughput because this particular VNF requires frequent memory accesses.

\fakepar{Correlation with latency}
As shown in Tables~\ref{table:latency_load_stimulus}, and~\ref{table:latency_resource_stimulus},  cache-related features show the strongest correlation with the latency. 
However, latency correlations are weaker overall because latency depends on high-level events such as buffer overflow and bandwidth saturation.

%latency generally has relatively weaker correlations with the hardware features compared to throughput. This is because latency, measured as the round-trip time of PTP packets, is heavily influenced by high-level system events such as buffer overflow and bandwidth saturation, which low-level features cannot easily capture. 

\begin{observation}
Some hardware features trend closely with the input traffic patterns across different service topologies, making them strong candidates for intermediate variables between input traffic and output KPIs. Additionally, these features capture the unique execution characteristics of individual VNFs, providing valuable insights into their behavior.
\end{observation}

%% file: paper/Sec4-SystemDesignNew.tex
%!TEX root = ../main.tex

\section{System design}
\label{sec:design}

We now present DRST, a framework that enables accurate performance inference, bottleneck diagnosis and drift-aware adaptive retraining.  We first outline the design principles, then present the end-to-end ML pipeline. Finally, we detail the MLOps implementation that supports deployment and monitoring.

%\todo{There is some redundancy with previous work, we can compress. - System design: I tried to clarify some concepts. We don't do fully hardware reading, the final part is always software. }

%\todo{We still need historical data, but how do we store it?}

%\todo{We still rely on stimuli to infer the system state, but it is not clear how we move from controlled environment to a real (or realistic) case}

\subsection{Design principles} ~\label{sec:arch}

\begin{figure*}[!tb]
\centering
\includegraphics[width=\textwidth]{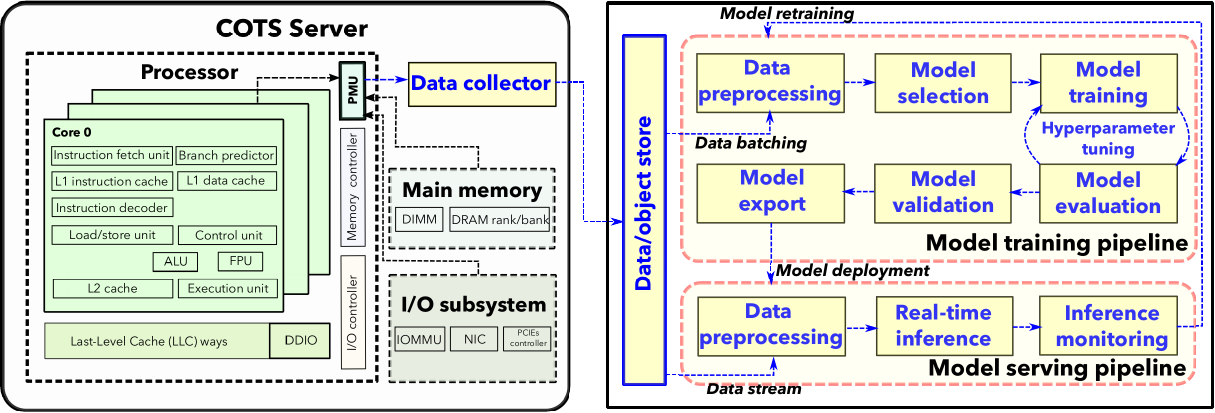}
\caption{Overall DRST architecture for performance prediction in high-speed softwarized networks.}
\label{fig:arch}
\end{figure*}

Based on the discussion of the prior works in Sec.~\ref {sec:related}, our architecture should respect the following design considerations. 
First, it must remain \textit{general} enough to deliver accurate and efficient performance impairment predictions for both individual VNFs and ground-up network services with limited deployment knowledge. Second, given the ever-increasing complexity of modern networks, it must be \textit{lightweight} and easy to deploy with minimal engineering exertions. As part of the network management subsystem colocated with VNF execution, it should be \textit{noninvasive} and introduce a negligible impact on the software data plane's normal operations and traffic. Finally, it must be \textit{fast} to enable real-time predictions using the available data. Fig.~\ref{fig:arch} illustrates our approach, the workflow has two steps: (i) data collection and  (ii) statistical learning.

{As an alternative to direct traffic-level measurements, we extract low-level hardware features from the shared network infrastructure.} %\footnote{\textbf{Leo Note} \emph{per-packet data collection overhead question} Modified.}
Although the internals of COTS servers are extremely complex, modern systems commonly offer various toolsets for performance monitoring, namely the Performance Monitoring Units (PMUs). It exposes hundreds of micro-architectural events that we can record in production~\cite{bakhvalov2020performance}.
This approach has several advantages.
First, PMU counters are always available and can be readily collected via standard profiling interfaces. High-level safeguard measures (e.g., encryption, private enclaves) do not hinder the collection of these low-level features. 
%Second, compared to per-packet measurement, collecting low-level features only happens at the hardware registers and thus causes much less intervention on the data path. Many low-level features are already available in the system registers, and our approach merely reuses them. 
Second, they impose negligible overhead even at Gigabit Ethernet. %\footnote{\textbf{Leo Note:} \emph{The features are stored in hardware, but reading is still via software} clarify this point.} 
This point is especially crucial in high-speed networks since even slight noise can cause noticeable performance losses~\cite{zhang2019comparing}. 
Third, our approach does not mandate an in-depth understanding of the target NFV system internals, such as the service, the management \& operation (MANO) plane, and the implementation details of (third-party) VNFs. Operators therefore avoid code instrumentation and extra integration work.

%\subsection{End-to-end ML pipeline}
%\label{mlpipeline}

Our end-to-end workflow comprises two parallel pipelines: a \textbf{Init-training} pipeline and an \textbf{Online-serving} pipeline. The design follows two  principles: (i) MLOps for reproducible and automated lifecycle management;  (ii) XAI for built-in interpretability.

\subsection{Init-training pipeline}

The offline training pipeline builds an initial model using historical data collected under controlled yet diverse configurations. Once trained, the model is integrated into the online-serving pipeline, which receives live feature streams via Kafka to support real-time KPI inference and forecasting. 

Note that the online pipeline also monitors input distributions to detect data drift and trigger lightweight model self-recovery mechanisms when necessary.
\subsubsection{Data preprocessing and feature selection}

Our data collection leverages multiple system layers to collect performance-relevant signals from both the system and application perspectives. At the NFV platform layer, we track hardware-level counters, including CPU cycles, instruction completions, cache operations, and PCIe bandwidth. In parallel, we use Intel PCM to track memory bandwidth, latency, and NUMA behavior. End-to-end throughput and latency are measured at the endpoints of the service chain. To emphasize the prediction task, we created 16 packet processing scenarios based on realistic SFC topologies and operational patterns, which encompass a wide range of memory and CPU behavior.

Differences in format and timing across tools make synchronization a non-trivial engineering task. We launch all monitoring tools under a synchronized trigger and align using timestamps. Next, the raw traces are passed through a preprocessing module, which handles vectorization, normalization, unified tagging, and cleanup.  The feature selection is based on the analysis in Section~\ref{sec:correlated_metrics}, which reveals a strong empirical correlation with target KPIs while filtering out redundant dimensions that could obscure the learning process.

%\begin{itemize}

 %  \item Data preprocessing: 
  % The pre-processing stage performs three actions: (1) Feature parsing \& normalization: convert %\texttt{perf} event tuples into a unified numeric vector and apply standardization. (2) Scenario tagging \& merge: attach <topology, stimulus> labels and combine trace segments into scenario-level samples. In total, we collected 16 scenarios to emulate the real network environment. (3) Quality filtering: drop incomplete rows and align PCM timestamps with \textit{perf} samples to ensure temporal consistency. 
%
 %  \item Feature selection: Following the feature relevance analysis in Section~\ref{sec:correlated_metrics}, we retain system-level metrics exhibiting strong correlation with target KPIs (throughput and latency), while minimizing inter-feature redundancy.
%\end{itemize}

\subsubsection{Scenarios construction}

\fakepar{Scenario-driven KPI estimation}
We represent each scenario as
\begin{align}
  \mathcal{E}&= (\mathcal{S},\texttt{Stim})
  \label{eq:scenario}
\end{align}
where $ \mathcal{S}= (v,\texttt{Topo})$, and $\texttt{Stim}= (\texttt{type},\mathbf{p})$.  $v=\{v_1,\dots,v_k\}$ is the ordered set of VNFs,  
\texttt{Topo} encodes their interconnection,  
\texttt{type} denotes the stimulus category, and  
$\mathbf{p}$ is a parameter vector (e.g., input-rate, CPU throttle).  
This abstraction underpins all evaluation scenarios (Section~\ref{sec:exp}).

%\fakepar{Scenario-Driven KPI Estimation}
%To train baseline model, we model system behavior using a tuple-based abstraction. Each scenario is defined as $\mathcal{E} = (\mathcal{S}, \texttt{Stim}) $, where $\mathcal{S} $ denotes the deployed SFC, and $\texttt{Stim}$ specifies the injected stimulus. The SFC is given by $\mathcal{S} = (v, \texttt{Topo}) $, where $v = \{v_1^S, v_2^S, \dots, v_k^S\}$ is the ordered set of VNFs, and \texttt{Topo} encodes their interconnection structure. The stimulus is defined as $\texttt{Stim} = (\texttt{type}, \mathbf{p}) $, where \texttt{type} indicates the stimulus category, and $\mathbf{p}$ is a parameter vector describing its intensity (e.g., traffic input rate, CPU saturation level). This abstraction forms the basis for the scenario construction in our evaluation (see Section~\ref{sec:exp}).

Given a scenario $\mathcal{E}$, the inference function $f^{\mathcal{E}}$ maps system inputs to KPI metrics such as throughput and latency. For multi-step forecasting, we denote the model output as $[\hat{y}_{t+1}, \dots, \hat{y}_{t+H}] = f^{\mathcal{E}}(X_{t-N+1:t})$, where $X_{t-N+1:t}$ represents the input window.

\fakepar{Inference engine: non-sequential mapping}
To infer instantaneous KPIs (e.g., throughput, latency) from hardware-level features, we define the inference engine $f^\mathcal{E}$ as a parametric mapping:
\begin{align}
  y = f^\mathcal{E}(\bm{X})
\end{align}
where $\bm{X}$ is the current observation vector, and $y$ is the target KPI. We implement a multi-layer perceptron (MLP) as the default inference model $f^\mathcal{E}$, as it offers the best trade-off between resource usage and latency as proofed in Section~\ref{sec:exp}.
We keep the inference component modular, making it easy to replace with other backends (e.g., tree ensembles) if deployment needs evolve.

The MLP is trained via regularized log-likelihood:
\begin{align}
\hat J(\bm{\theta}) = \frac{1}{n} \sum_{i=1}^n L(\bm{X}_i, y_i, \bm{\theta}) + \frac{\alpha}{2}\|\bm{\theta}\|^2_2
\end{align}
where $\alpha$ controls L2 regularization to prevent overfitting under high dynamic range, we use Adam for gradient-based minimization. 

\fakepar{Forecasting engine: sequential prediction}
\label{sec:lstm}
To enable proactive control, the forecasting task aims to predict future KPI values $[y_{t+1}, \cdots, y_{t+H}]$ from a history of observations $[\bm{X}_{t}, \cdots, \bm{X}_{t-N+1}]$. Our evaluation covers seven models from four major families: regression, tree-based, LSTM, and Transformer. Here, we focus on two  LSTM-based designs due to their strong temporal modeling capabilities.

\begin{itemize}
    \item Standard LSTM: This model directly maps the input window to multiple future steps in parallel:
\begin{align}
[\hat{y}_{t+1}, \dots, \hat{y}_{t+H}] = f^{\mathcal{E}}(\bm{X}_{t-N+1:t})
\label{eq:std_lstm}
\end{align}
\item {DirREC-LSTM}~\cite{han2019review} recursively feeds its predictions back as inputs to improve long-horizon accuracy:
\begin{align*}
\hat{y}_{t+h} =
\begin{cases}
f^{\mathcal{E}}_1(\bm{X}_{t-N+1:t}), & h = 1 \\
f^{\mathcal{E}}_h(\hat{y}_{t+1:t+h-1}, \bm{X}_{t-N+1:t}), & 2 \leq h \leq H
\end{cases}
\label{eq:dirrec}
\end{align*}
The recursive design enhances prediction fidelity in long-tail scenarios, but increases inference latency due to its sequential nature.
\end{itemize}
Note that DirREC-LSTM is more accurate in complex traffic scenarios. However, its runtime cost is higher, which is impractical for real-time inference on constrained hardware. We therefore adopt standard LSTM as the default model. The module remains replaceable with alternative learners.

We employed a grid search strategy for hyperparameter tuning due to its simplicity and interpretability. Due to space constraints, not all experimental results are included here.  For instance, we found that the number of hidden layers and nodes per layer had less influence on overall accuracy compared to other factors such as the optimizer and batch size in MLP-based models.  We thus limited the number of hidden layers in the MLP to be $\leq4$, and the number of nodes per layer to be $\leq64$.  In parallel, we varied the number of stacked LSTM layers (1–3) and tested hidden state dimensions ranging from 32 to 128. We also examined different sequence lengths, since the input window size shows a tradeoff between accuracy and efficiency. Compared to MLP, LSTM performance was more sensitive to batch size and sequence length. 

Finally, the validated model, along with its configuration and metadata, is packaged in a versioned internal registry for integration into the serving pipeline.

%\subsubsection{Hyperparameter Search, Validation, and Export}
%\begin{itemize}

 %   \item Hyperparameter Search: We adopt a modular tuning stage that supports interchangeable strategies such as grid search, random sampling, and Bayesian optimization, all implemented via a unified interface. In this work, we used grid search for offline training.

   % \item Final Training: The selected model is retrained on the full dataset using the best hyperparameters to maximize generalization and statistical coverage.

   % \item Model Validation: This stage confirms the quality of the trained model through its accuracy, F1-score, and inference latency on a held-out test set. The results serve  as a reference baseline for online prediction.
%
  %  \item Model Export: The validated model, along with its configuration and metadata, is packaged and stored in a versioned internal registry for integration into the serving pipeline.
%\end{itemize}

\subsubsection{Model Interpretability via SHAP}
To improve the interpretability of the learned model $f^E$, we employ SHAP~\cite{lundberg2017unified} to understand how the input features $\bm{X}$ contribute to the predicted output $y$. By integrating SHAP explanations with domain knowledge, we can extract actionable insights into how hardware-level metrics influence KPIs. SHAP assigns each feature a SHapley value, capturing its marginal contribution. Formally, for a prediction $f^E(\mathbf{X})$, the contribution of each feature $\mathbf{X}_j$ is  
{ \small\begin{align*}
\phi_j\!\! = \!\!\sum_{D \subseteq \{1, \dots, d\} \setminus \{j\}} \frac{|D|! (d - |D| - 1)!}{d!} \left[ f^E(\bm{X}_{D \cup \{j\}}) - f^E(\bm{X}_D) \right],
\end{align*}}
where $d$ is the total number of features, and  $\bm{X}_D$  is the input restricted to subset $D$. The term $f^E(\bm{X}_{D \cup \{j\}}) - f^E(\bm{X}_D)$ denotes the marginal contribution of feature  $X_j $ when added to subset $D$. Note that the sum of all feature contributions equals the model's output, i.e., $ f^E(\bm{X}) = \phi_0 + \sum_{j=1}^d \phi_j$, and $\phi_0$ is the baseline predictive performance without any feature permutation.

   \begin{figure}
    \centering
    \includegraphics[width=0.95\linewidth]{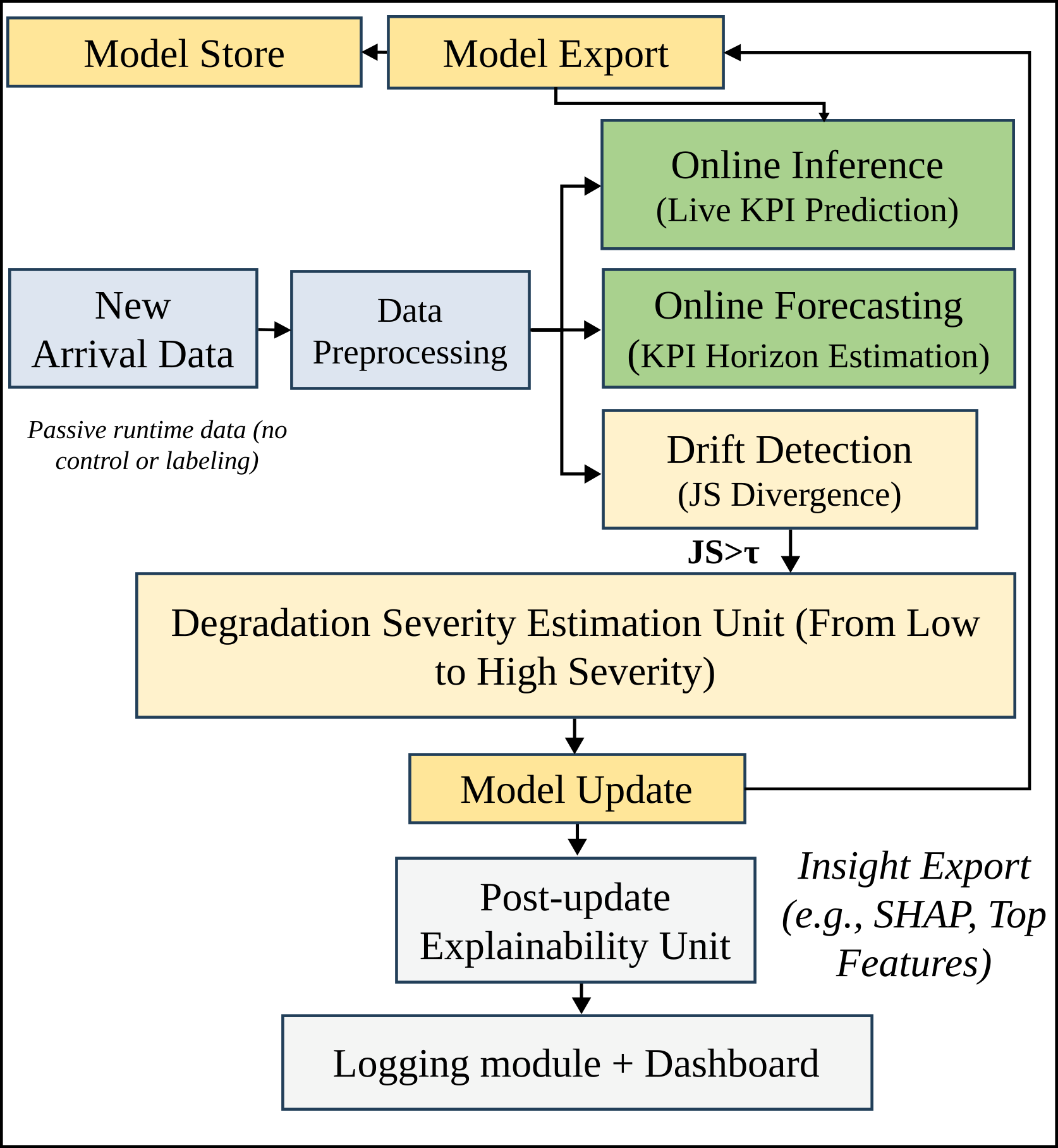}
    \caption{ Model serving pipeline. All components operate on live, unlabeled data to support passive inference, drift detection, and automated model updates.
}
    \label{fig:serve_pipelines}
\end{figure}
 \subsection{Model serving pipeline} 
\label{sec:serv_pipeline}

The serving pipeline performs passive, real-time inference and adaptation without relying on controlled stimuli and explicit labels. It consumes live runtime features and operates continuously after deployment. %\footnote{\textbf{Leo notes:} \emph{We still rely on stimuli to infer the system state, but it is not clear how we move from controlled environment to a real (or realistic) case} Updated context.} 
As shown in Fig.~\ref{fig:serve_pipelines}, it comprises modular components for data handling, inference, drift analysis, model update, and system monitoring. %, thereby supporting adaptive and explainable operation. %To preserve real-time inference responsiveness, we decouple the inference and monitoring paths. While incoming data streams are first ingested through a message queue, latency-sensitive inference operations consume directly with aggressive polling, whereas drift detection operates asynchronously with buffered batches.

%\item {Data Acquisition:} Each input sample is automatically constructed by aligning multiple time-synchronized monitoring streams collected from various components (e.g., VNFs, memory, PCIe, and packet I/O statistics). The aligned values are assembled into fixed-order feature vectors according to a predefined schema that matches the model input format. This preprocessing is fully automated and does not involve any form of ground-truth labeling. The resulting feature vectors are concurrently dispatched to the online inference and drift detection modules for parallel execution.
%To ensure consistency with the training pipeline, each incoming sample is preprocessed with feature normalization and encoding. The processed data is then concurrently dispatched to the online inference and drift detection modules for parallel processing.
\vspace{1em}
\noindent\textbf{Data handling~~~}At runtime, monitoring streams from VNFs collect data from multiple hardware features, including memory access, PCIe patterns, and CPU consumption linked to packet I/O. Such data are time-aligned and assembled into fixed-order feature vectors matching the model input schema. Then, the resulting vectors are streamed via Kafka and dispatched in parallel to the inference, forecasting, and drift detection modules for concurrent execution, as shown in the "Data" blocks of Fig.\ref{fig:serve_pipelines}.

%\footnote{\textbf{Leo Note} \emph{We should probably give an example of how one inference is made from the real system. Besides the formulas in Sec.4, we can show an algorithm (high-level) that shows how we interact between data, operations, inference, and retraining. This is done in Fig.6 or Fig.7, but it is not very clear to me. Automated labeling is important: how do we do this, if we do not control the environment? This still goes a bit against the claim "without prior knowledge"? } To clarify: during inference we don't actually need any ground-truth labels or automated labeling, since the system only uses runtime features and unsupervised signals like JS divergence to detect drift and trigger updates.  I modify the first para to make things more clear. }
\vspace{1em}
\noindent\textbf{Inference and forecasting~~~} The two following blocks represent the online processes of inference and forecasting. The inference engine consumes each incoming feature vector and produces real-time KPI predictions using the most recent model generated from the training or adapted pipeline. The forecasting engine runs in parallel with inference to project short-horizon KPI trajectories based on recent input windows. This enables proactive detection of emerging anomalies or performance shifts before they appear in observed KPIs.

%\item Drift Detection: The same preprocessed data is concurrently analyzed by a drift detection module based on Jensen-Shannon (JS) divergence, quantifying distributional shifts relative to historical reference windows.

\vspace{1em}
\noindent\textbf{Drift analysis~~~} To understand if a drift is occurring, we employ a drift detection module, which continuously monitors distribution changes in the input features using a sliding-window approach. At each step, it compares the current window of $M$ (e.g., 100) recent samples against a historical reference window using Jensen-Shannon (JS) divergence. This unsupervised method quantifies input drift over time without requiring ground-truth labels and serves as the trigger signal for downstream model updates.

\begin{algorithm}[tb]
\SetAlgoLined
\KwIn{Monitoring features $\bm{X}_t$ arriving at time $t$}
\KwOut{Predicted KPI $\hat{y}_t$, possible model update}
$\mathcal{W}_{\text{curr}} \gets$ sliding window of recent $M$ samples\;
\While{system is running}{
    Receive new feature vector $\bm{X}_t$\;
    Push $\bm{X}_t$ to Kafka queue\;
    Predict $\hat{y}_t \gets f_{\text{current}}(\bm{X}_t)$\;
    Forecast $\hat{y}_{t+1:t+H} \gets f_{\text{forecast}}(\bm{X}_{t-N+1:t})$\;
    Update $\mathcal{W}_{\text{curr}}$ with $\bm{X}_t$\;
    \If{size$(\mathcal{W}_{\text{curr}}) = M$}{
        Compute JS divergence $D_{\text{JS}}$ against $\mathcal{W}_{\text{ref}}$\;
        \If{$D_{\text{JS}} > \delta$}{
            severity $\gets$ estimate\_degradation($D_{\text{JS}}$)\;
            $f_{\text{new}} \gets$ retrain\_model(severity)\;
            $f_{\text{current}} \gets f_{\text{new}}$\;
            $\mathcal{W}_{\text{ref}} \gets \mathcal{W}_{\text{curr}}$\;
        }
    }
    Log outputs and metrics\;
}
\caption{Online Inference and Drift-triggered Update Loop}
\label{alg:drift_loop}
\end{algorithm}

%\item Degradation Estimation and Update Triggering: When the JS divergence exceeds a configurable threshold, the system estimates the degradation severity and selects a corresponding update strategy from three predefined levels of complexity (e.g., adjusting the learning rate, tuning the architecture, or expanding the search space).

A second module estimates the degradation severity and decides whether to trigger an update:
when the observed JS divergence exceeds a configurable threshold, the system estimates the severity of the drift and dynamically selects an update strategy. Update levels are categorized by complexity, ranging from lightweight re-optimization (e.g., learning rate tuning) to full retraining with extended search space. 

%\item Model Update and Export: The selected retraining procedure is launched, and the updated model is exported to replace the current one. All inference outputs, divergence logs, and model metadata are tracked for downstream monitoring.

\vspace{1em}
\noindent\textbf{Model Update and Export~~~}  Once triggered, the selected retraining procedure is executed automatically, producing a new model version packaged with its configuration and metadata. The updated model replaces the active one in the serving pipeline, and all related outputs—including predictions, drift scores, and version history are logged for traceability and future analysis.

\vspace{1em}
\noindent\textbf{Monitoring and Explainability~~~} Runtime metrics, including per-sample latency and throughput, are collected with Prometheus and shown on a dashboard for live monitoring and alerting. Prediction results, update triggers, and SHAP-based explanations are visualized through a lightweight UI dashboard for human-in-the-loop interpretability.

%\item \textbf{Monitoring and Explainability:}   Runtime performance metrics, such as per-sample latency and throughput, are collected via Prometheus and visualized through a system dashboard for continuous monitoring. Prediction outputs, drift events, and SHAP-based feature attributions are also exposed through a lightweight UI to support operator insight and system transparency.

%Figure~\ref{fig:serve_pipelines} illustrates the serving pipeline architecture and data flow. 
To complement the component-level view, Algorithm~\ref{alg:drift_loop} provides a runtime perspective, showing how the system performs online inference  in a continuous loop.

\begin{figure}
    \centering
\includegraphics[width=1\linewidth]{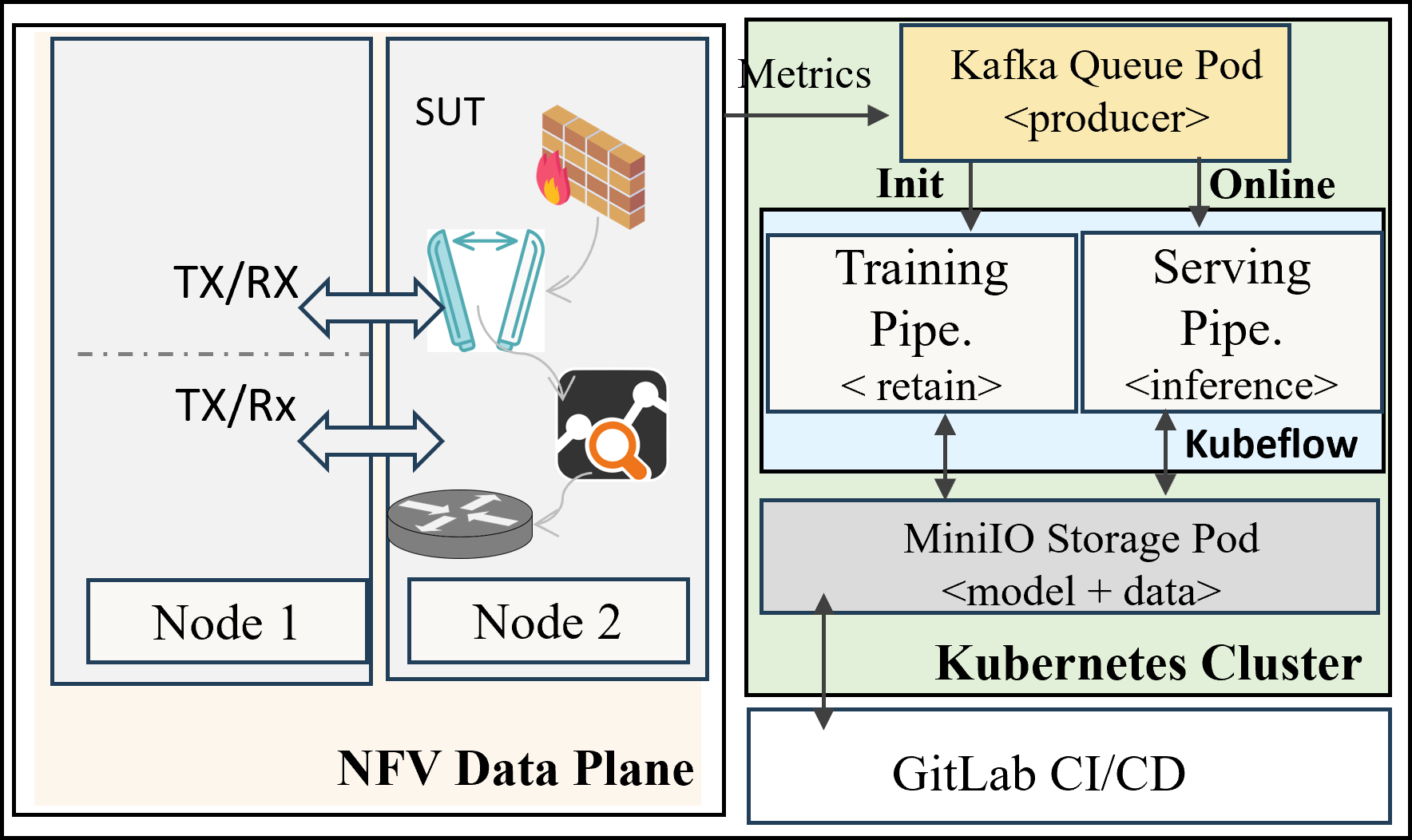}
    \caption{Infrastructure-level MLOps Implementation}
    \label{fig:Platform}
\end{figure}

\begin{table*}[tb]
\scriptsize
  \centering
  \setlength{\tabcolsep}{1pt}
  \begin{tabular}{|l|c|c|c|c|c|c|c|c|c|}
    \hline
    \multirow{2}{*}{\bf Server} & \multicolumn{3}{c|}{\bf Intel Xeon Processor} & \multirow{2}{*}{\bf Memory} & \bf NUMA & \multicolumn{2}{c|}{\bf Last-Level Cache} & \multicolumn{2}{c|}{\bf NIC} \\
    \cline{2-4} \cline{7-10}
     & \bf CPU Model & \bf Frequency & \bf \#Cores & & \bf Sockets & \bf Size & \bf \#Ways & \bf Brand & \bf Speed \\
    \hline
    Skylake & Silver 4210R & 2.4 GHz  & 20 & 128 GiB & 2 & 27.5 MiB & 11 & Broadcom\textsuperscript{\textregistered} BCM5741X & $4\times25$~Gbps \\
    Broadwell & E5-2640 v4 & 2.4 GHz & 20 & 64 GiB & 1& 25.6 MiB & 20 & Mellanox\textsuperscript{\textregistered} MT27710 CX-4 & $1\times40$~Gbps\\
    Haswell & E5-2667 v3 & 2.6 GHz & 20 & 64 GiB & 2 & 19.7 MiB & 20 & Intel\textsuperscript{\textregistered} 82599ES 10-Gigabit & $4\times10$ Gbps \\
    \hline
  \end{tabular}
  \caption{Testbed specifications}
  \label{tab:servers}
\end{table*}

\begin{table*}[!tb]
\scriptsize
\centering
\setlength{\tabcolsep}{3pt}
\begin{tabular}{|l|l|l|}
\hline
\textbf{Category} & \textbf{Component} & \textbf{Version / Notes} \\
\hline
\multirow{3}{*}{System \& HW} 
  & CPU / Memory & AMD EPYC 12 vCPU, 48 GB ECC \\
  & Disk / fio & 800 GiB SATA SSD, R/W $\approx$ 400–450 MB/s, IOPS $\approx$1.6–1.7k \\
\hline
\multirow{4}{*}{Container Stack} 
  & Docker Engine & v28.0.1 \\
  & Kubernetes & v1.28.15 (Kind v0.27.0, kubectl v1.32.2) \\
  & Helm / Kustomize & Helm v3.17.2, Kustomize v5.5.0 \\
  & Prometheus / Python & Prometheus 2.x, Python 3.11.11 \\
\hline
\multirow{5}{*}{MLOps Stack} 
  & Kubeflow Pipelines & v2.4.0, SDK v2.5.0 (DSL v2) \\
  & Katib / KServe & Katib v0.18.0-rc.0, KServe v0.14.1 \\
  & Training Operator & v1-5170a36 (TFJob / PTJob) \\
  & Kafka & Bitnami 3.7.1-debian-12-r4 \\
  & MinIO / MySQL & MinIO RELEASE.2019-08-14, MySQL 8.0.29 \\
\hline
\end{tabular}
\caption{Execution environment  for containerized pipeline experiments}
\label{tab:mlops-env}
\end{table*}

\subsection{Deployment on a Kubernetes MLOps system}
\label{sec:imp}

The proposed system is deployed on a \emph{Kubernetes cluster} and implemented via a modular MLOps stack centered on \emph{Kubeflow} Pipelines. All ML components, including the training and serving pipelines, are containerized as independent services and orchestrated as DAGs. 

We leverage \emph{Kafka} as the real-time message bus. Feature vectors exported by VNFs are published to Kafka topics and consumed in parallel by three services: the MLP-based inference engine, the LSTM forecaster engine, and the JS divergence detector engine. These components are decoupled and horizontally scalable. The JS divergence is computed over sliding windows to quantify input drift. Once the divergence exceeds the threshold, the system launches a lightweight model update without interrupting service.

We use \emph{MinIO}, a shared object store, as the internal model registry. It stores historical datasets, inference logs, model artifacts, and version metadata.  Model updates are published to MinIO, and the serving engines regularly pull the latest versions. This separation of storage and execution supports rollback deployment strategies. We use a \emph{GitLab CI/CD} for version control, training reproducibility, and pipeline tracking.

Figure~\ref{fig:Platform} summarizes the infrastructure-level implementation of the proposed MLOps architecture. On the left, the NFV data plane consists of two NUMA nodes that host the system under test (SUT), which is responsible for generating and capturing system-level metrics. %%% LL: It is not Packet-level, this word is misleading, we do CPU-level (so I changed to system level
These metrics are streamed into the control plane via a Kafka queue. On the right, the Kubernetes-based MLOps control stack includes modular training and serving pipelines orchestrated by Kubeflow, a shared object store (MinIO), and a CI/CD layer for version control and deployment. The decoupled training and serving workflows allow independent lifecycle management, while MinIO serves as a persistent backend for both model artifacts and streaming data. The infrastructure-level implementation DAG is illustrated in Fig.~\ref{fig:app-platform} (see ~\ref{app:infra}).

%% file: paper/Sec5-Experiments.tex
%!TEX root = ../main.tex

\section{Experimental evaluation}
\label{sec:exp}

\subsection{Testbed environment}\label{sec:testbed}

\fakepar{Hardware settings}
To validate the applicability of our framework, we conduct experiments on three types of COTS servers with diverse hardware components, as illustrated in Tab.~\ref{tab:servers}. As indicated by the server names, the servers have processors spanning three generations of Intel CPU microarchitectures, i.e., Haswell, Broadwell, and Skylake. 
%with two Intel\textsuperscript{\textregistered} Xeon CPUs E5-2660 v3 @ 2.60GHz with a three-level cache of 64K/256K/25600K. Each processor consists of 10 physical cores attached to a NUMA node. 
To minimize interference, all the cores used for our tests are isolated from the kernel scheduler (via {\em isolcpus}) with hyper-threading and turbo-boost disabled.

\fakepar{Software Settings} We use MoonGen~\cite{emmerich2015moongen}, a high-speed packet processing engine, to generate traffic and measure the end-to-end KPIs, specifically throughput and latency. At runtime, we collect the low-level hardware features of individual VNFs using {\textit{perf}}~\cite{perf} and {\textit{PCM}}~\cite{pcm}, two standard system profiling tools interfacing with the system PMUs at a configurable frequency (e.g., 200/500/1000 ms). The sampling frequency reflects a trade-off between temporal resolution and system overhead. Shorter intervals capture fine-grained performance fluctuations (e.g., bursty contention or latency microbursts) but incur higher profiling overhead that could interfere with VNF execution. 
%\todoA{precision: why different frenquency? }
The low-level features of each VNF can be collected via pre-assigned execution identifiers (e.g., process, thread, or function IDs). Note that other standard profiling tools~\cite{bakhvalov2020performance} can also serve the same purpose. 

\begin{table*}[!t]
\scriptsize
\centering
\begin{tabular}{lcccccc}
\toprule
\multirow{2}{*}{\textbf{Model}} &
\multicolumn{3}{c}{\textbf{Seen trace (train/test split)}} &
\multicolumn{2}{c}{\textbf{Unseen trace}} &
\multirow{2}{*}{\textbf{Latency (ms/sample)}}\\
\cmidrule(lr){2-4}\cmidrule(lr){5-6}
 & \textbf{$R^{2}$} & \textbf{MAE} & \textbf{Acc@5\,\%$_{\log}$} & \textbf{MAE} & \textbf{Acc@5\,\%$_{\log}$} & \\ \midrule
Linear              & {0.98} & \textbf{183.52} & 97.02 & \textbf{163.93} & 92.98 & 0.030\\
SVR                 & 0.26 & 2122.68 & 46.00 & 2064.17 & 47.09 & 0.373\\
Decision Tree        & 0.96 & 309.46 & 96.78 & 281.43 & 87.92 & 0.030\\
Random Forest        & 0.97 & 286.33 & 97.34 & 258.70 & 88.42 & 0.030\\
Gradient Boosting    & 0.96 & 429.67 & 95.10 & 399.22 & 86.86 & \textbf{0.001}\\
XGBoost             & 0.96 & 429.66 & 94.92 & 403.11 & 86.35 & 0.042\\
\textbf{MLP}        & \textbf{0.99} & 257.78 & \textbf{98.28} & 211.41 & \textbf{96.81} & 0.652\\
\bottomrule
\end{tabular}
\caption{Inference engine selection comparison (\emph{Best value per column in bold}).}
\label{tab:inf_perf}
\end{table*}

\fakepar{Network service deployment}\label{sec:sfc}
We deploy the VNFs on NUMA node 0 using OpenNetVM (ONVM)~\cite{zhang2016opennetvm}, which provides flexible service composition and high-speed packet steering. The instantiated VNFs operate as bare-metal processes and can be purposely connected to form an intended network service. Each VNF runs in busy-polling mode, which monopolizes a CPU core (with 100\% usage).  
Note that ONVM is selected because of its impressive performance and accessibility; our approach should apply to other prevalent NFV frameworks (e.g., ClickOS~\cite{martins2014clickos} or E2~\cite{palkar2015e2}).  The evaluation spans a range of service chains and runtime topologies, ensuring that the observed behavior is not tied to a specific deployment and that our conclusions remain topology-agnostic. %\todoA{Are all data put together in a big collected data set? If so, we must explain that the system is topology-agnostic.}

\fakepar{Workload generation}
Performance issues generally arise due to overwhelming loads and insufficient resources~\cite{aghasaryan2017stimulus}. 
We consider two basic workload generation schemes to expose latent performance issues: {\em load stimulus} and {\em resource stimulus}. The former composes the input traffic with special patterns to contrive load contentions, while the latter perturbs the resource shares of individual VNFs to fabricate resource contentions. To better control the imposed contentions, we employ competitor processes to expose and analyze the impact of resource contentions. The competitors are based on {\em stress-ng}~\cite{stress-ng}, a standard stress-testing tool capable of generating bogus operations at different subsystem components, including CPU pipeline, multi-level caches, and memory.  For example, CPU contention can be created by pinning parasite competitors to a VNF's worker core. The cache contentions can be generated by thrashing existing lines. The memory bandwidth contention can be induced by injecting I/O requests. We can even generate multiple contentions by calculatedly assigning the competitors.

\fakepar{MLOps execution stack}
We deploy the model inference and adaptation workflow on a containerized Kubernetes cluster orchestrated with Kubeflow Pipelines and Argo. Inference is served via KServe, and hyperparameter tuning is managed by Katib. Each model component is containerized and deployed declaratively. Input data is streamed through Kafka, and model artifacts are stored in MinIO using S3-compatible access. System metrics, such as resource usage, inference latency, and update frequency, are collected via Prometheus and visualized using Grafana dashboards and Python scripts (see Table~\ref{tab:mlops-env}).  \footnote{We re-stream previously sampled KPI traces via Kafka with 1:1 timing replay, faithfully reproducing the original real-time sampling intervals.}  
To ensure that storage access does not bottleneck inference or training, we benchmark the disk performance using \texttt{fio}. The measured sequential read and write throughputs are 442 MB/s and 446 MB/s, respectively. The average latency per I/O operation is approximately $300~\mu$s on the MinIO array.

%\fakepar{Metric Collection and Measurement}
%We report standard system and inference metrics to evaluate pipeline responsiveness and efficiency. Throughput is defined as the number of successfully processed inference requests per second, measured at the consumer side of Kafka. Latency includes the full inference path, from Kafka message ingestion to prediction output and Kafka result emission, with timestamps inserted before and after model execution. For pipeline-level diagnostics, we extract task durations and scheduling overhead directly from Argo Workflow logs, using per-step start/end timestamps.

\begin{table*}[ht]
\centering \scriptsize
\begin{tabular}{lcccccccccccccccc}
\toprule
\multirow{2}{*}{\textbf{Model}} & 
\multicolumn{2}{c}{\textbf{R$^2$}} &
\multicolumn{2}{c}{\textbf{MAE}} &
\multicolumn{2}{c}{\textbf{Latency (ms)}} &
\multicolumn{2}{c}{\textbf{Acc@t+1}} &
\multicolumn{2}{c}{\textbf{Acc@t+2}} &
\multicolumn{2}{c}{\textbf{Acc@t+3}} &
\multicolumn{2}{c}{\textbf{Acc@t+4}} &
\multicolumn{2}{c}{\textbf{Acc@t+5}} \\
\cmidrule(r){2-3} \cmidrule(r){4-5} \cmidrule(r){6-7}
\cmidrule(r){8-9} \cmidrule(r){10-11} \cmidrule(r){12-13}
\cmidrule(r){14-15} \cmidrule(r){16-17}
 & \cellcolor{blue!10}A & \cellcolor{orange!10}B 
 & \cellcolor{blue!10}A & \cellcolor{orange!10}B 
 & \cellcolor{blue!10}A & \cellcolor{orange!10}B 
 & \cellcolor{blue!10}A & \cellcolor{orange!10}B 
 & \cellcolor{blue!10}A & \cellcolor{orange!10}B 
 & \cellcolor{blue!10}A & \cellcolor{orange!10}B 
 & \cellcolor{blue!10}A & \cellcolor{orange!10}B 
 & \cellcolor{blue!10}A & \cellcolor{orange!10}B \\
\midrule
LSTM               & 0.95 & 0.61 & 346.6 & 591.7 & \textbf{2.04} & \textbf{2.86} & \textbf{0.99} & {0.96} & 0.96 & \textbf{0.96} & 0.93 & \textbf{0.93} & 0.90 & 0.86 & 0.87 & 0.81 \\
DirectLSTM         & 0.96 & \textbf{0.68} & 316.2 & \textbf{521.5} & 324.3 & 424.2 & \textbf{0.99} & \textbf{0.97} & 0.96 & \textbf{0.96} & 0.94 & \textbf{0.93} & \textbf{0.92} & \textbf{0.89} & \textbf{0.90} & \textbf{0.82} \\
Ridge  & 0.98 & \textbf{0.73} & 216.4 & 569.8 & {1.04} & \textbf{0.21} & 0.99 & {0.95} & 0.94 & {0.86} & 0.96 & 0.81 & 0.78 & 0.75 & 0.63 & 0.60 \\
Random Forest      & 0.95 & 0.65 & 280.0 & 605.0 & 70.00 & 65.00 & 0.96 & 0.93 & 0.94 & 0.91 & 0.92 & 0.87 & 0.90 & 0.83 & 0.89 & 0.82 \\
XGBoost            & \textbf{0.99} & 0.65 & \textbf{94.1} & 629.2 & 16.42 & 58.93 & 0.98 & \textbf{0.94} & 0.98 & 0.93 & 0.98 & 0.90 & 0.86 & 0.84 & 0.84 & 0.82 \\
TransformerLight   & 0.90 & 0.54 & 621.0 & 740.2 & 2.11 & 3.43 & 0.96 & 0.89 & 0.96 & 0.87 & 0.93 & 0.84 & 0.93 & 0.83 & 0.88 & 0.81 \\
Transformer        & 0.94 & 0.60 & 450.3 & 718.7 & 2.56 & 5.86 & 0.98 & 0.97 & 0.98 & 0.89 & \textbf{0.98} & 0.92 & 0.88 & 0.86 & 0.84 & 0.83 \\
\bottomrule
\end{tabular}
\caption{Comparison of forecasting models under two runtime traffic patterns: \cellcolor{blue!10}A (periodic trend), \cellcolor{orange!10}B (stage-random fluctuations).}
\label{tab:forecast_combined_full}
\end{table*}

\subsection{Model selection and generalization analysis}
\label{sec:model-selection}
This section evaluates candidate models before integrating them into the serving pipeline.
\begin{table*}[!t]
\centering
\scriptsize
\begin{tabular}{llcccc}
\toprule
\textbf{Topology} & \textbf{Scenario Type} & \textbf{Acc@5\% (Mean $\pm$ Std)} & \textbf{MAPE (Mean $\pm$ Std)} & \textbf{Latency (ms)} & \textbf{Throughput (samples/s)} \\
\midrule
Single VNF        & Load Stimulus (Regular)     & 98.2\% $\pm$ 1.4\% & 2.1\% $\pm$ 0.7\% & 0.74 & 1351 \\
\midrule
\multirow{4}{*}{Linear SFC} 
                  & Load Stimulus               & 97.4\% $\pm$ 1.9\% & 2.5\% $\pm$ 0.6\% & 0.76 & 1315 \\
                  & Random Traffic Stimulus     & 92.8\% $\pm$ 2.7\% & 4.1\% $\pm$ 1.2\% & 0.78 & 1282 \\
                  & Resource Contention         & 83.6\% $\pm$ 3.1\% & 6.3\% $\pm$ 1.7\% & 0.75 & 1333 \\
                  & Intervention                & 72.4\% $\pm$ 3.5\% & 7.9\% $\pm$ 2.0\% & 0.77 & 1310 \\
\midrule
\multirow{3}{*}{DAG1 SFC}
                  & Load Stimulus               & 95.1\% $\pm$ 1.8\% & 3.1\% $\pm$ 1.0\% & 0.73 & 1369 \\
                  & Random Traffic Stimulus     & 90.2\% $\pm$ 2.5\% & 5.2\% $\pm$ 1.3\% & 0.74 & 1351 \\
                  & Resource Contention         & 82.5\% $\pm$ 2.9\% & 7.3\% $\pm$ 1.9\% & 0.76 & 1315 \\
\midrule
\multirow{3}{*}{DAG2 SFC}
                  & Load Stimulus               & 93.6\% $\pm$ 2.0\% & 3.7\% $\pm$ 1.1\% & 0.75 & 1333 \\
                  & Resource Contention         & 79.8\% $\pm$ 3.6\% & 8.2\% $\pm$ 2.1\% & 0.78 & 1282 \\
                  & Intervention                & 67.9\% $\pm$ 3.8\% & 8.8\% $\pm$ 2.3\% & 0.76 & 1315 \\
\bottomrule
\end{tabular}
\caption{MLP-based throughput inference performance across runtime scenarios. Accuracy is reported using  Acc@5\% and MAPE, averaged over 10 i.i.d. test runs (170 samples each). Inference latency is measured per sample, excluding Kafka overhead. Throughput is measured via Prometheus-based monitoring.}
\label{tab:appendix-throughput}
\end{table*}

\subsubsection{Inference engine selection}
To compare accuracy and serving costs, we evaluate eight regression models, as shown in Table~\ref{tab:inf_perf}.
All models are trained on a representative workload trace comprising \num{9788} samples (80 MB) and evaluated on an 80/20 split.  A second trace with \emph{random-stage} traffic is used to test generalization.  Inference latency is measured as the wall-clock time per sample on a single CPU core.  Accuracy is reported by the coefficient of determination ($R^{2}$), the mean-absolute error (MAE), and the log-space accuracy within 5\% \footnote{Due to the inherent limitations of software traffic generators~\cite{emmerich2015moongen}, the rate gaps in high-speed regimes make it difficult to estimate the exact rates accurately. According to our micro-benchmarks, $5\%$ is a realistic error.} (Acc@5\,\%$_{\log}$), the latter mitigating scale imbalance in the target variable. 

%We have found that most models, except SVR, perform similarly well on the seen data ($R^2 \ge 0.96$). SVR struggles likely due to sensitivity to feature scaling and large target variations.
%In terms of generalizability, the MLP maintains the highest log-Acc@5\% (96.8\%) on the unseen trace, with only a modest MAE increase. %In contrast, tree models show notable drops, suggesting weaker generalization.
%Ensemble trees (RF, XGB) deliver sub-millisecond inference, while the MLP incurs a higher cost due to dense-matrix computations. Despite a higher latency (0.65\,ms vs 0.05\,ms), MLP remains sub-millisecond and batch-amenable within the NFV data plane system's sampling intervals.% (typically 200–1000ms, as configured in \textit{perf} and \textit{PCM}).

We have found that most models, except SVR, perform similarly well on the seen data ($R^2 \ge 0.96$). SVR struggles likely due to sensitivity to feature scaling and large target variations.  Among all tested models, MLP achieves the best generalization, reaching a log-Acc@5\% of 96.8\% on unseen traces. On the other hand, ensemble trees (RF, XGB) deliver sub-millisecond inference, which results in the lowest inference latency. MLP incurs a slightly higher cost due to dense matrix computations, as  0.65\,ms vs 0.05\,ms. However, it remains sub-millisecond and batch-amenable with the NFV data-plane system sampling intervals.  Given its strong generalization performance, the MLP may reduce the frequency of model retraining or updates. We therefore adopt the MLP as the default model for performance inference.

%We choose MLP as the default offline model due to its generalization ability over acceptable latency gains.

\subsubsection{Forecasting engine selection}
The forecasting engine periodically predicts short-horizon throughput to support proactive scaling. We benchmark several models designed to capture temporal dynamics, including \textit{Ridge}, \textit{XGBoost}, \textit{standard and direct LSTM}, and \textit{Transformer-based} models. All models are evaluated with an identical input sequence length of 10 steps and a forecast horizon of 5 steps.

We evaluate two runtime patterns: \textit{Pattern A}, representing periodic traffic, and \textit{Pattern B}, characterized by stage-wise bursts with abrupt plateaus and shifts. Table~\ref{tab:forecast_combined_full} reports accuracy ($R^2$, MAE), short-horizon reliability (Acc@t+$k$), and per-sample latency.

%Our results show that tree-based models (Random Forest, XGBoost) exhibit low latency and strong accuracy under regular workloads, but suffer notable performance drops in the presence of irregular fluctuations. DirREC-LSTM achieves better generalization across both regimes but incurs higher inference cost due to cumulative dependence across forecast steps. In contrast, the standard LSTM offers a favorable trade-off: it maintains high step-wise accuracy (Acc@t+1 to Acc@t+5) and consistent performance across both traffic patterns, while remaining computationally lightweight. Given its forecasting accuracy and operational efficiency, LSTM is adopted as the default forecasting module in our pipeline. %Ridge remains available as a
%fallback option in compute-constrained deployments, despite its limitations on complex sequence patterns.

Our results show that tree-based models (Random Forest, XGBoost) exhibit low latency and strong accuracy under regular workloads, but suffer large performance drops under irregular traffic fluctuations.  Second, DirREC-LSTM achieves the best generalization across scenarios. However, it incurs higher inference cost due to cumulative dependence across forecast steps—each future step depends on the output of the previous one. By comparison, the standard LSTM strikes a practical balance: it maintains high prediction accuracy while performing all predictions in a single forward pass. Considering both accuracy and runtime cost,  we select the standard LSTM as a performance forecasting engine.

\subsection{Online inference performance}
We evaluate inference performance over three representative SFC topologies (defined in Figure~\ref{fig:sfcs} and described in Section~\ref{sec:testbed}), each composed of 5--6 VNFs such as Firewall, nDPI-stat, and Payload-scan.

\subsubsection{Throughput inference} \label{sec: thr_pred}

\begin{figure*}[h]
\centering
\begin{subfigure}[b]{0.33\textwidth}
    \centering
    \includegraphics[width=2.2in]{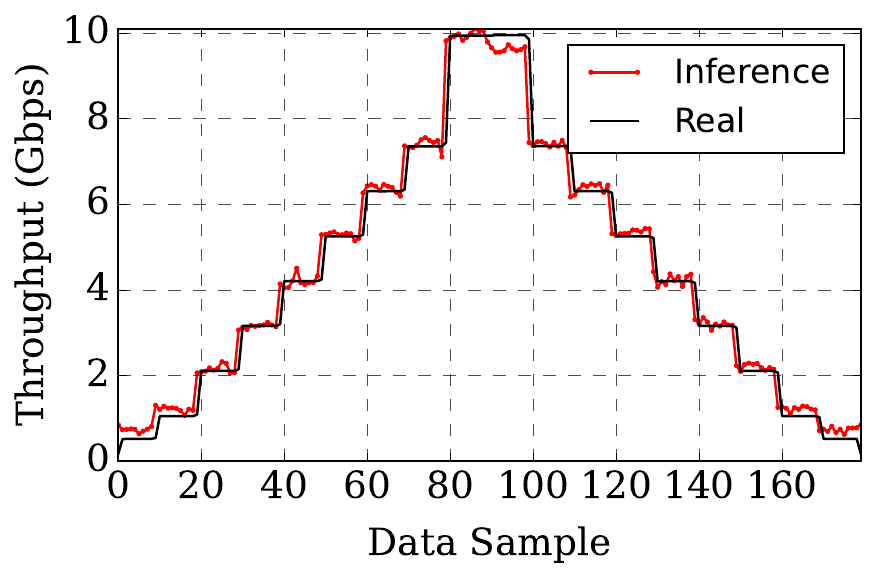}
    \caption{Linear (regular rate)}
    \label{fig:inference_1}
\end{subfigure}%
\begin{subfigure}[b]{0.33\textwidth}
    \centering
    \includegraphics[width=2.2in]{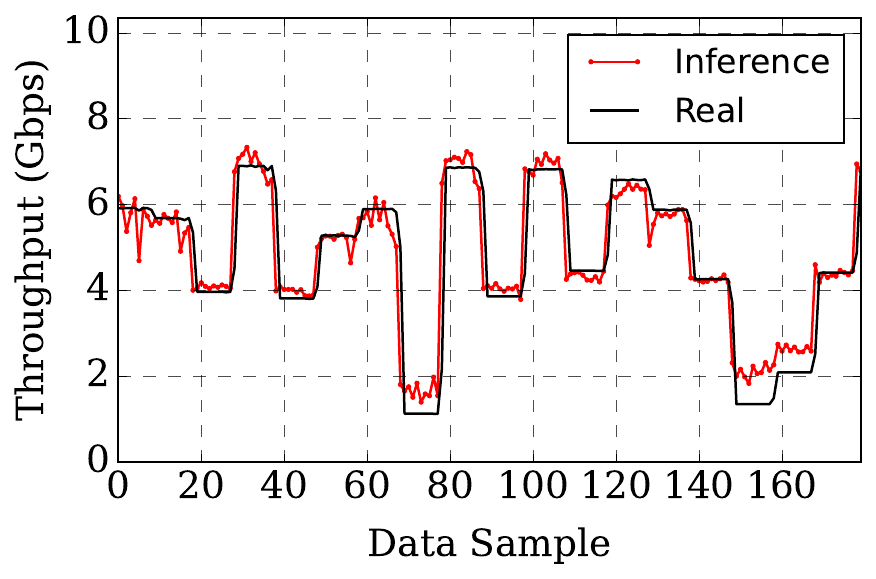}
    \caption{Linear (random rate)}
    \label{fig:inference_2}
\end{subfigure}
\begin{subfigure}[b]{0.33\textwidth}
    \centering
    \includegraphics[width=2.2in]{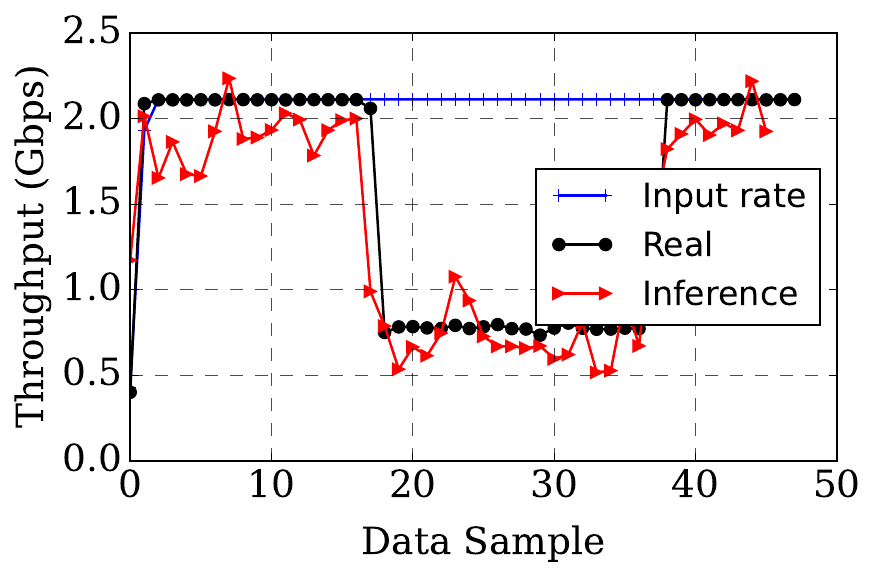}
    \caption{Linear (res. stim.)}
    \label{fig:inference_3}
\end{subfigure}
\caption{Throughput inference under different workload generation}
\end{figure*}

% Accuracy is evaluated using the mean absolute percentage error  with a $5\%$ tolerance. 
%Under load stimulus, our model achieves impressive overall accuracy: $98\%$ for the regular rate and $92\%$ for the random rate, as illustrated by the examples in Fig.~\ref{fig:inference_1} and \ref{fig:inference_2}. Under resource stimulus, throughput inference becomes more intricate due to diversiform contentions from the CPU, caches, and memory buses. Still, our model's accuracy remains commendable at $83\%$, as illustrated in Fig.~\ref{fig:inference_3}. 
\begin{figure}[h]
    \centering
\includegraphics[width=.95\linewidth]{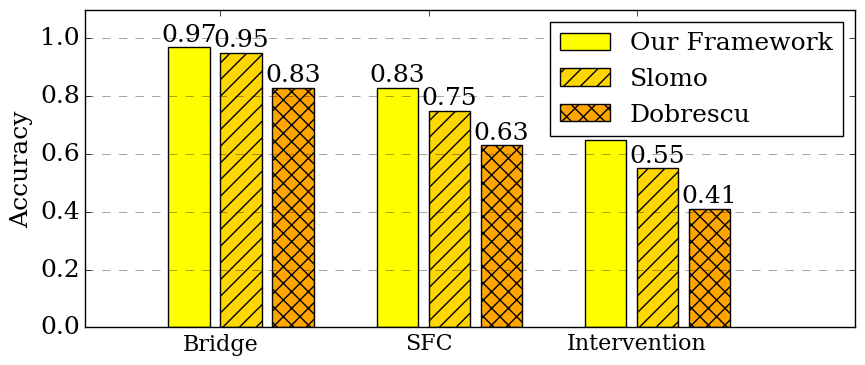}
    \caption{Throughput inference comparison}
    \label{fig:throughput-comparison}
\end{figure}
\fakepar{Accuracy across stimuli}
We evaluate throughput inference under realistic serving conditions by streaming time-aligned traces through Kafka and invoking per-sample inference via KServe, consistent with the production-serving pipeline. While the evaluation utilizes replayed traces for reproducibility and traceability, the inference engine operates in an online manner, processing one input at a time, within containerized serving endpoints. For each runtime scenario, 10 independent test runs are conducted, each consisting of 170 samples. Accuracy is reported as the mean and standard deviation of log-space Acc@5\% and MAPE across 10 runs. %We use a logarithmic error scale to account for the wide dynamic range of throughput values.

Under load stimulus, our model achieves impressive overall accuracy: $98\%$ for the regular rate and $92\%$ for the random rate, as illustrated by the examples in Fig.~\ref{fig:inference_1} and \ref{fig:inference_2}. Under resource stimulus, throughput inference becomes more intricate due to diverse contentions from the CPU, caches, and memory buses. Still, our model's accuracy remains commendable at $83\%$, as in Fig.~\ref{fig:inference_3}. All per-scenario results are further summarized in Table~\ref{tab:appendix-throughput}.%, which lists the inference performance across various deployment topologies and runtime stimuli. 

\fakepar{Comparison with SoTA} Prior works on NFV throughput inference primarily attribute performance degradation to memory subsystem bottlenecks. Dobrescu et al.~\cite{dobrescu2012toward} propose a linear model based on cache access rates, assuming additive contention effects across traffic flows. SLOMO~\cite{manousis2020contention} improves estimation accuracy by jointly modeling cache and memory bandwidth contention using gradient boosting regression, but is evaluated only in single-VNF setups and lacks robustness under non-linear contention patterns. We compare our model against these two approaches under three scenarios as in Fig~\ref{fig:throughput-comparison}:   \textbf{(1)} a singleton VNF (ONVM bridge) under mixed stimuli,   \textbf{(2)} a linear SFC under resource contention, and   \textbf{(3)} the same SFC subject to both load and resource interference.   To enable a fair and deployment-relevant comparison, we re-implement both Dobrescu and SLOMO models and integrate them into the same containerized pipeline used for our inference engine.

Our feature set differs slightly from SLOMO by excluding L2 cache and L3 occupancy features. This discrepancy does not affect the overall accuracy as we have already included abounding highly pertinent cache features. In scenario 1, our model can accurately predict throughput, with a mean accuracy of $97\%$, which outperforms SLOMO's $95\%$ and Dobrescu's $83\%$; 
Overall, our model increases the prediction accuracy concerning Dobrescu and SLOMO by $18\%$ and $7\%$ on average, respectively, and reduces the average prediction error by $70\%$ and $37\%$.
Note that under the most challenging scenario 3, our model's accuracy is $65\%$, which might further benefit from data enhancements.

% Our approach improves over these works along three key dimensions:
%\begin{itemize}
 % \item \textit{Scenario coverage.} While previous work focuses on isolated VNFs or synthetic workloads, we build and evaluate a full testbed spanning 11 real-world configurations.%—including linear and DAG-shaped SFCs—with diverse interference sources (load, resource, and intervention).
 %   \item \textit{Deployment realism.} Unlike prior offline evaluations, our inference engine runs in a live, containerized environment. Models are deployed into a Kubernetes-based pipeline, enabling real-time per-sample prediction with observability through Prometheus.
 %   \item \textit{Accuracy comparison.} In Figure~\ref{fig:throughput-comparison}, we compare inference accuracy in three scenarios of increasing complexity. Even in the most stressed setting (SFC under dual interference), our model maintains 65\% accuracy—compared to 41\% and 56\% for Dobrescu and SLOMO, respectively.
%\end{itemize}

%Note that our feature set differs slightly from SLOMO by excluding L2 cache and L3 occupancy features. This discrepancy does not affect the overall accuracy as we have already included abounding highly pertinent cache features.

\subsubsection{Latency inference}
\label{sec: predic latency}

Latency inference is harder due to drastic state transitions under congestion (Sec.~\ref{sec:correlated_metrics}). Here, the latency is measured as the round-trip delay of probe packets generated by MoonGen at 1~Gbps, with a maximum wait time of 120~ms per probe. These probe packets are processed concurrently with regular service traffic, offering a realistic estimate of packet processing delay.

%Under normal traffic and resource conditions, the measured latency remains stable below $50~\mu$s and exhibits minimal variance. In such regimes, latency inference is of limited operational significance. However, we design a controlled workload where the traffic rate increases linearly from 0 to the line rate (10~Gbps), and then decreases back to 0. Despite the smooth traffic pattern, the observed latency does not grow linearly. It remains low (<$50~\mu$s) for most of the duration, but exhibits sharp spikes only when approaching saturation. This indicates that latency jumps abruptly once critical thresholds are exceeded, making it challenging to predict using continuous traffic or resource metrics alone.

\begin{figure}[!tb]
  \centering
  \begin{subfigure}[b]{0.23\textwidth}
    \includegraphics[width=\textwidth]{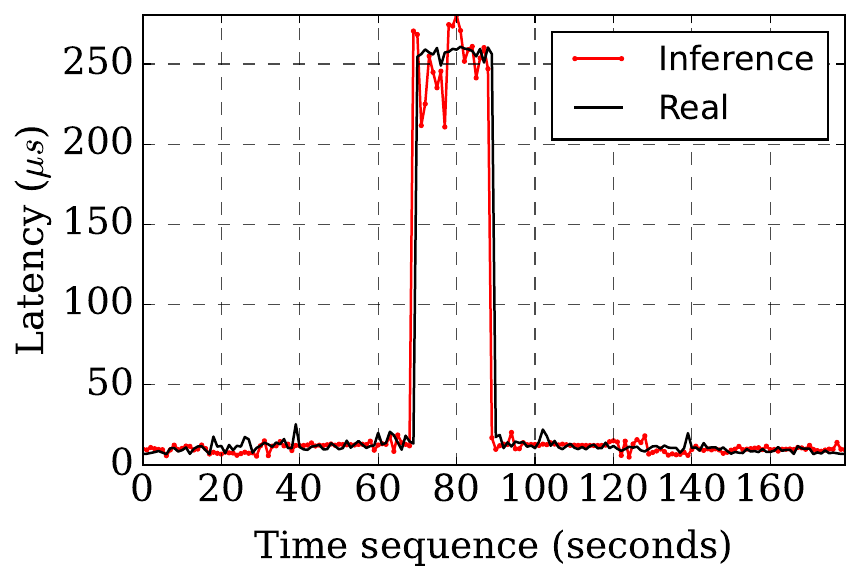}
    \caption{Linear}
    \label{fig:latency_a}
  \end{subfigure}
  \hfill
  \begin{subfigure}[b]{0.23\textwidth}
  \includegraphics[width=\textwidth]{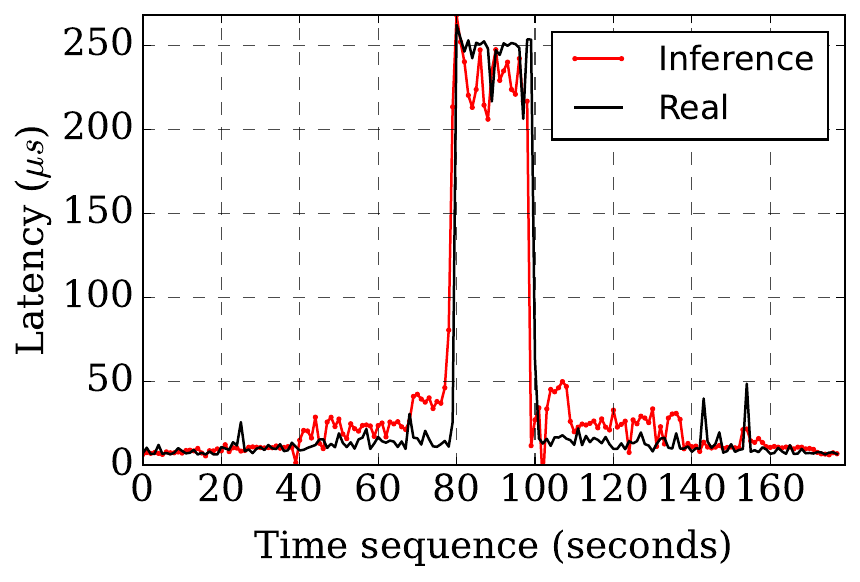}
    \caption{DAG-1}
    \label{fig:latency_b}
  \end{subfigure}  
   \begin{subfigure}
     [b]{0.49\textwidth}
  \includegraphics[width=\textwidth]{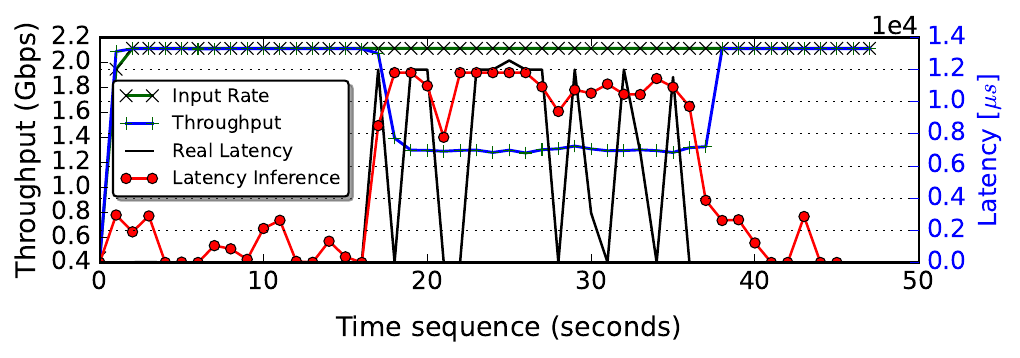}
  \caption{Resource stimulus}
  \label{fig:latency_c}
  \end{subfigure}
  \caption{Latency inference under three scenarios}
  \label{fig:latency_prediction}
\end{figure}

\fakepar{Prediction accuracy} 
 Fig.~\ref{fig:latency_a} and Fig.~\ref{fig:latency_b} show its performance on linear and DAG-1 topologies under increasing load. The model achieves 86\% and 79\% accuracy, respectively, except in the high-throughput regime (80–100s), where congestion and queue build-up cause RTT to oscillate.

\fakepar{Stress under resource contention}
Fig.~\ref{fig:latency_c} shows the inference results under resource stimulus. Severe contention between 15s and 35s cuts throughput by up to 40\%, triggering packet losses and erratic delays. During such periods, latency becomes fundamentally unpredictable, but our model can still approximate its envelope and detect high-latency transitions. Current methods are unsuitable for predicting the exact (artificial) latency under severe network congestion, as sporadic packet losses make the RTT hard to quantify. However, our model can still approximate the expected values with prior knowledge of the maximum latency and predict abnormal service
latency and congestion periods.

%\subsubsection{Latency inference}
%\label{sec: predic latency}
%Latency inference is intrinsically complicated due to more drastic state transitions under congestion, and pairwise correlation analysis is inadequate to uncover the complicated relationships, as discussed in Sec.~\ref{sec:correlated_metrics}. 

\subsubsection{Explainability and feature sensitivity}

To facilitate interpretation, we design an XAI unit pod by applying a SHAP explainer to assess feature importance in KPI inference. To reduce the high computational complexity ($\mathcal{O}(n \cdot d^2 \cdot T_{\text{inference}})$), we first employ gradient-based sensitivity analysis to pre-select the most relevant features. This reduces the feature set size (e.g., from 42 to 10), cutting SHAP runtime from 2400~s to 170~s on a 100-sample dataset.

\begin{figure}[!tb]
    \centering
\includegraphics[width=.85\linewidth]{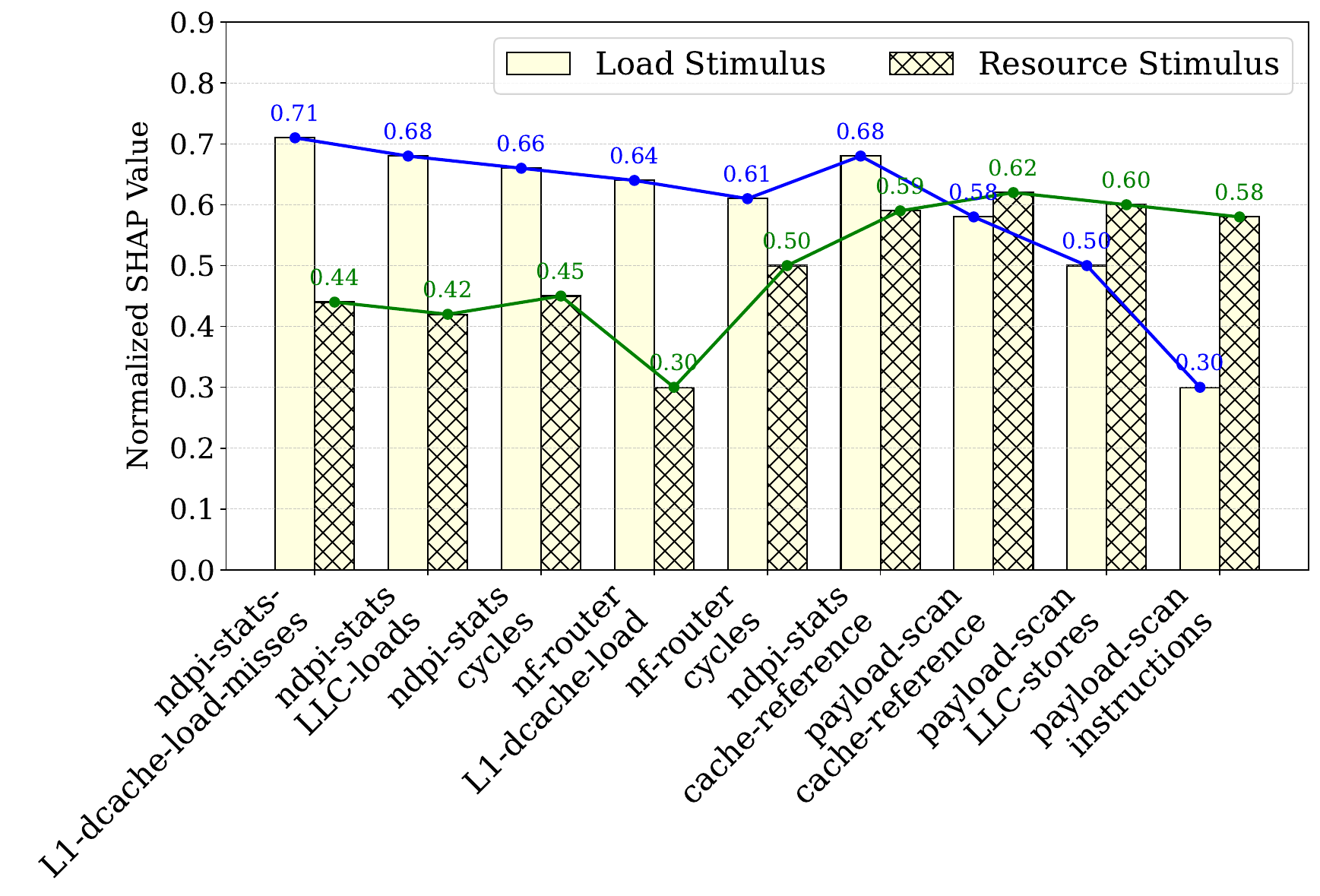}
    \caption{XAI analysis for throughput inference. }
    \label{fig:XAI_resource}
\end{figure}

\fakepar{SHAP for throughput}  Fig.~\ref{fig:XAI_resource} shows the normalized SHAP values for the top features in throughput inference.   As expected, cache-related features dominate. However,  we found that \texttt{*\_cycles} features also show high attribution, which may contribute to the unstable CPU frequency scaling. Under \textbf{load stimulus}, \texttt{L1-dcache-load-misses} becomes the main contributor. This suggests that heavy traffic increases pressure on L1 caches, causing potential pipeline stalls. In the end,  under \textbf{resource stimulus}, features like \texttt{*\_cache-references} and \texttt{*\_instructions} stand out. This indicates last-level cache contention and potential DMA inefficiencies when resources are constrained.

\begin{figure}[!tb]
    \centering
\includegraphics[width=.85\linewidth]{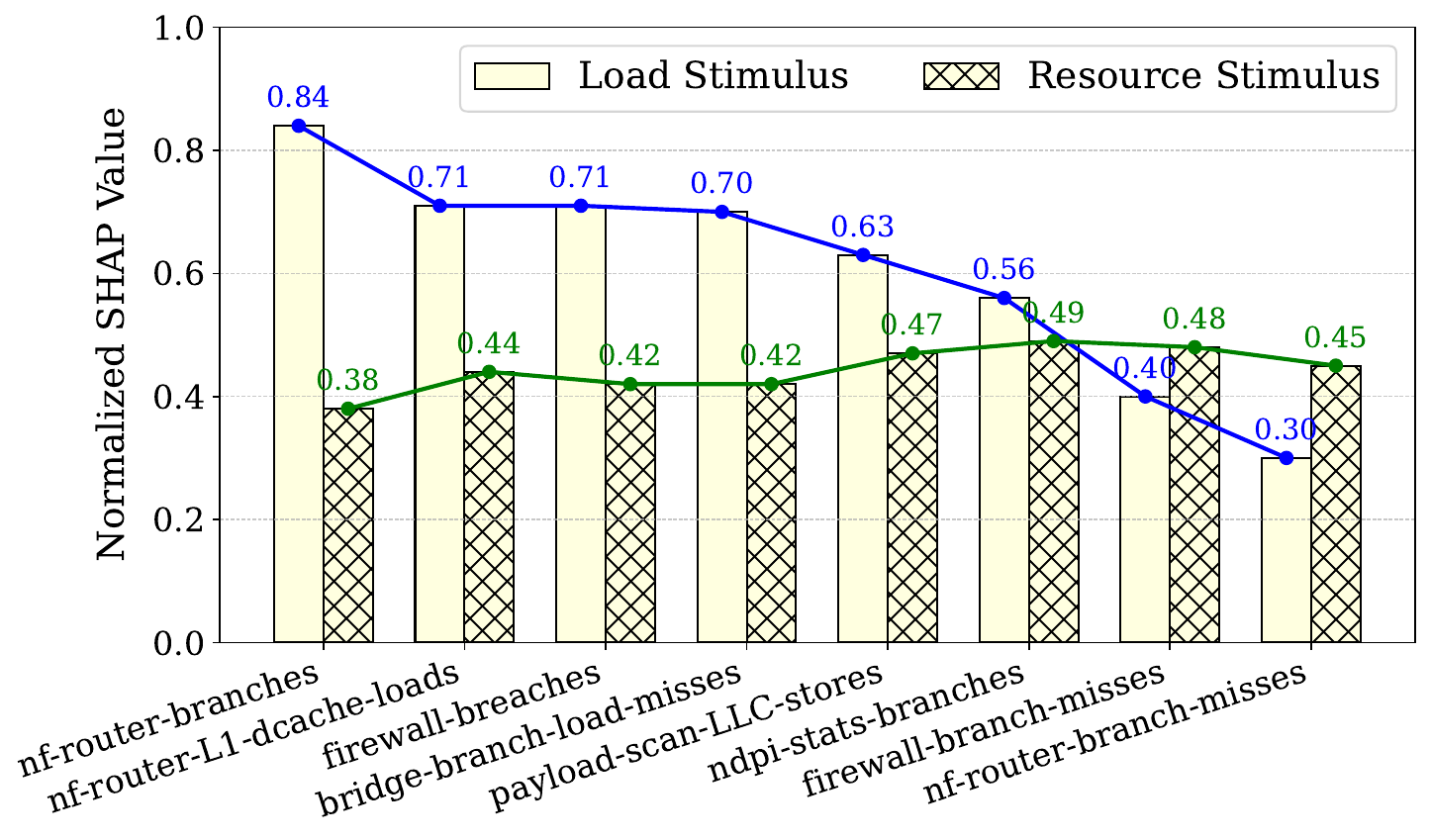}
    \caption{XAI analysis for latency inference. }
    \label{fig:xai_latency}
\end{figure}

\fakepar{SHAP for latency } Fig.~\ref{fig:xai_latency} shows the top SHAP-ranked features for latency inference. We summarize three key observations: 1) Branch-related features (e.g., \textit{*\_branch}, \textit{*\_branch\_miss}) dominate, highlighting the role of CPU branch prediction in latency behavior; 2) Under \textbf{load stimulus}, \textit{nf-router\_branches} is the top contributor, suggesting that peak traffic increases branching pressure in the router module; 3) Under \textbf{resource stimulus}, control-plane modules like \textit{ndpi-stats-branches} and \textit{firewall-branch-misses} become dominant, as limited resources amplify misprediction risks and delay packet processing.

\subsection{Throughput forecasting }
The forecasting module runs inside the real-time MLOps pipeline. Features vectors are streamed via Kafka, and predictions are computed directly on unseen inputs. The test data follows the same topology and traffic pattern (\texttt{linear-random-stage}), where the specific inputs used for evaluation are entirely disjoint from the training set. We consider a forecasting horizon of $H{=}12$ future steps. Each input is constructed from a 12-second window of performance metrics, collected at a 1-second sampling interval using Intel PCM. The forecasting targets correspond to the aggregated throughput (in Gbps) over the next 12 seconds. 

\begin{figure*}[!tb]
  \centering
  \begin{subfigure}[b]{0.32\textwidth}
    \includegraphics[width=\textwidth]{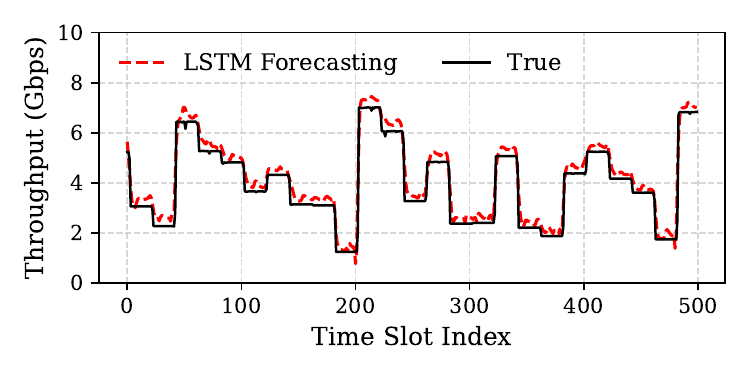}
    \caption{$t$+2}
    \label{fig:forecast_t2}
  \end{subfigure}
  \begin{subfigure}[b]{0.32\textwidth}
    \includegraphics[width=\textwidth]{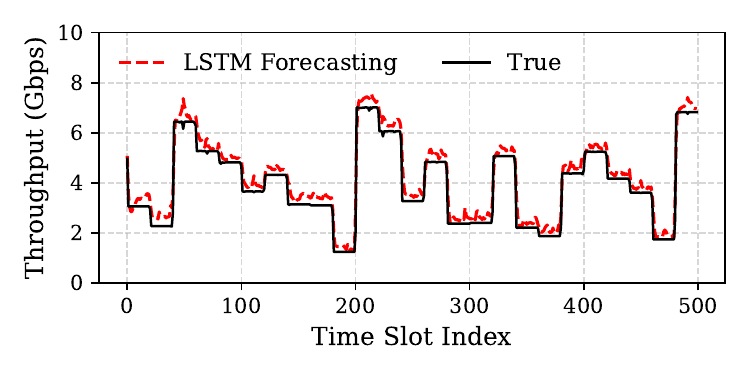}
    \caption{$t$+4}
    \label{fig:forecast_t4}
  \end{subfigure}
  \begin{subfigure}[b]{0.32\textwidth}
    \includegraphics[width=\textwidth]{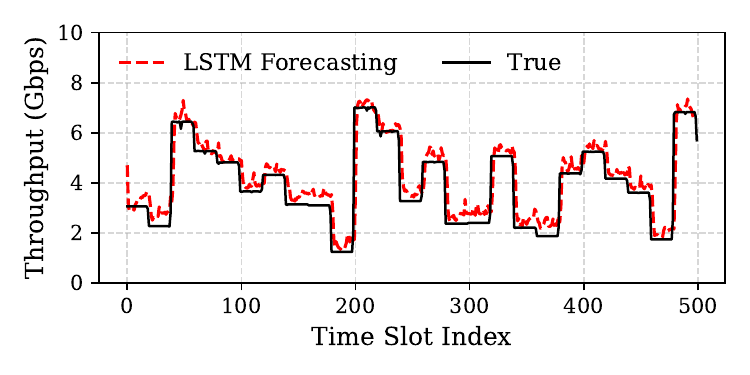}
    \caption{$t$+6}
    \label{fig:forecast_t6}
  \end{subfigure}
  \begin{subfigure}[b]{0.32\textwidth}
    \includegraphics[width=\textwidth]{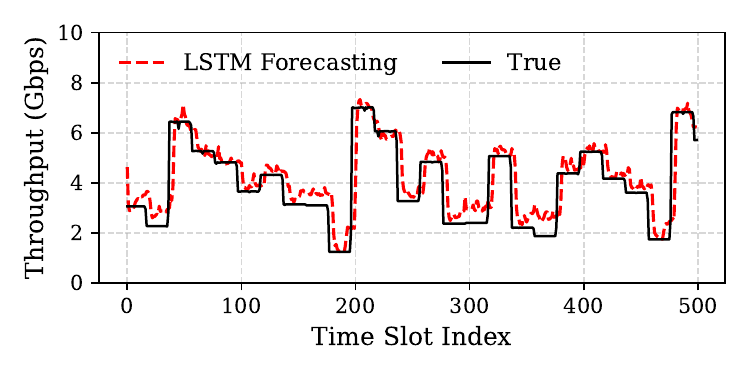}
    \caption{$t$+8}
    \label{fig:forecast_t8}
  \end{subfigure}
  \begin{subfigure}[b]{0.32\textwidth}
    \includegraphics[width=\textwidth]{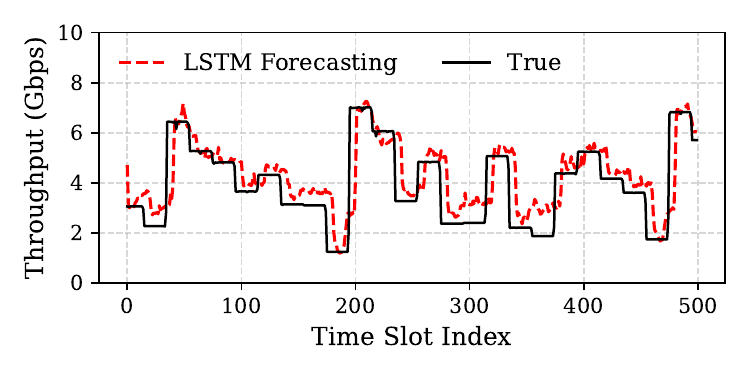}
    \caption{$t$+10}
    \label{fig:forecast_t10}
  \end{subfigure}
  \begin{subfigure}[b]{0.32\textwidth}
    \includegraphics[width=\textwidth]{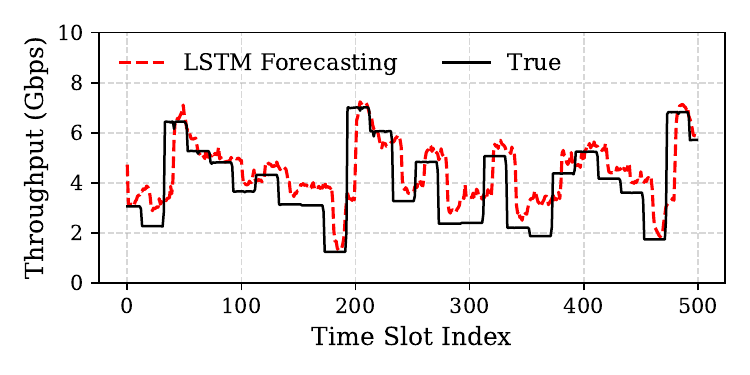}
    \caption{$t$+12}
    \label{fig:forecast_t12}
  \end{subfigure}
\caption{Online multi-step forecasting in the deployed MLOps pipeline. Each curve represents predictions for a specific forecast step ($t{+}2$ to $t{+}12$), computed from a single inference execution on unseen streamed input.}

  \label{fig:forecast_all}
\end{figure*}

As shown in Figure~\ref{fig:forecast_all}, the predicted throughput from the standard LSTM model is plotted against ground truth values for selected steps from $t{+}2$ to $t{+}12$.   As expected, accuracy decreases with longer horizons, dropping from $86.24\%$ at 2 seconds to $55.3\%$ at 12 seconds. 

\begin{table*}[!t]
\centering
\scriptsize

\begin{tabular}{llcccl}
\toprule
\textbf{Pipeline Stage} & \textbf{Module} & \textbf{Online} & \textbf{Runtime (s)} & \textbf{Trigger Frequency} & \textbf{Description} \\
\midrule
Data Handling       & Extraction       & $\checkmark$ & 0.01   & Per sample         & Stream raw feature vectors from Kafka \\
                    & Preprocessing         & $\checkmark$ & 0.03   & Per sample         & Normalize and transform input features \\
Inference           & Inference Engine      & $\checkmark$ & 0.01   & Per sample         & Predict current throughput (MLP) \\
                    & Forecasting Engine    & $\checkmark$ & 0.13   & Every 30s          & Predict near-future trend \\
Monitoring          & Drift Detection & $\checkmark$ & 0.01   & Every 10s          & Compute JS divergence between sliding windows \\
\midrule
Adaptation          & Severity-1 Handler    & $\times$      & 12.7   & On drift (low)     & Lightweight tuning (learning rate, batch size) \\
                    & Severity-2 Handler    & $\times$      & 47.3   & On drift (med)  & Moderate tuning (depth, activation, optimizer) \\
                    & Severity-K Handler    & $\times$      & 141.9  & On drift (severe)  & Full grid search \\
                    & XAI Unit              & $\times$      & 0.8    & On drift (med) & SHAP-based analysis to interpret model behavior \\
                     & Model Update          & $\times$      & <1.0   & Post-tuning        & Push retrained model to KServe and update endpoint \\
\bottomrule
\end{tabular}
\caption{Runtime profiling of key modules in the unified inference–adaptation pipeline (see Fig.~\ref{fig:app-platform}). Online modules are triggered per sample or periodically. Adaptation stages are invoked based on detected drift severity.}
\label{tab:runtime-breakdown}
\end{table*}
\subsection{System runtime and adaptation efficiency}
\label{sec:runtime}

In this subsection, we evaluate the runtime behavior of each module within the deployed pipeline. As summarized in Table~\ref{tab:runtime-breakdown}. Online components are either triggered per input sample or executed periodically, and adaptation stages are invoked based on the detected drift severity.  Specifically, the data-handling pod and inference engine operate on every incoming sample, and the forecasting engine is invoked every 30 seconds. Drift detection is triggered every 10 seconds,  using JS divergence between sliding windows as the drift indicator. We can see that the tasks are lightweight, with module runtime ranging from 0.01s (inference engine) to 0.03s (preprocessing), ensuring real-time performance under a per-second sampling interval.

\begin{table}%
\centering
\scriptsize
\begin{tabular}{llll}
\toprule
\textbf{Metric} & \textbf{Scope} & \textbf{ Value} \\
\midrule
Model Inference Latency\footnote{Measured at the KServe model server interface; differs from pipeline trace latency reported in Table~\ref{tab:runtime-breakdown}, which reflects internal runtime of the inference module only.} & Internal (MLP only) & 0.12 s \\
Pod Response Latency & End-to-end RTT & 0.02 s \\
Inference Throughput & Internal & 3180 samp/s \\
Cold Start Time & Startup Time (KServelogs) & 62.02 s \\
\bottomrule
\end{tabular}
\caption{End-to-end inference latency and deployment-time metrics.}
\label{tab:timeMetrics}
\end{table}

Drift detection is triggered periodically and completes within 9 ms per window, using JS divergence between feature distributions. Upon drift detection, the system invokes adaptation routines according to a severity-aware policy. Minor drifts (Severity-1) prompt quick tuning procedures (12.7s), moderate drifts (Severity-2) adjust model structure (47.3s), while severe drifts (Severity-K) initiate full hyperparameter search (up to 141.9s). We took $K=3$ in this experiment. Explainability modules (e.g., SHAP) are optionally enabled for in-depth analysis. Model updates are pushed online via a hot-swapping mechanism, with deployment latency under 1 second.

\begin{figure}
    \centering
    \includegraphics[width=1\linewidth]{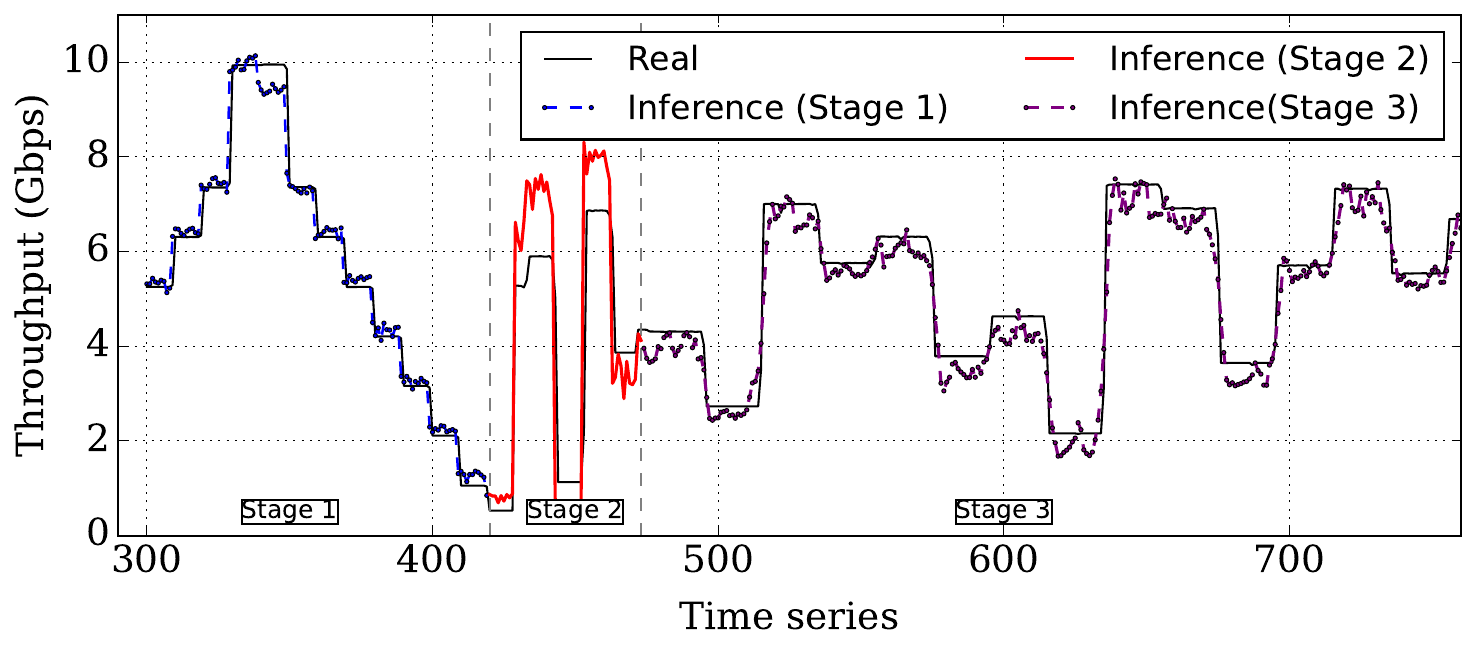}
    \caption{Drift detection and correction under load stimulus. The system transitions from Stage~1 (pre-drift) to Stage~2 (retraining), and resumes accurate inference in Stage~3 with the updated model.}
    \label{fig:drift-correction}
\end{figure}

To complement the runtime profiling and adaptation triggers summarized in Table~\ref{tab:runtime-breakdown}, we provide a time-series visualization of the drift detection and correction process in Fig.~\ref{fig:drift-correction}. The experiment simulates a resource-induced drift, where low-level features from a changed service topology (\texttt{DAG-1}) replace the initial regular load inputs (\texttt{bridge}). Upon detecting a significant distributional shift via JS divergence, the system triggers Severity-2 retraining. Within around 44 seconds, the new model is deployed, restoring prediction accuracy on the incoming data stream. Ta

%To further characterize system responsiveness and resource efficiency, we report detailed runtime statistics of two critical components. Table~\ref{tab:js-monitor-runtime} summarizes the monitoring performance of the JS Divergence module, showing consistently low CPU usage and memory footprint. The divergence computation completes within a few milliseconds per batch, with negligible Kafka lag, supporting high-frequency drift detection without impacting pipeline stability.

%Table~\ref{tab:timeMetrics} reports the inference and deployment-time metrics  measured across different system layers.  The model inference latency (0.12 s) includes tensor serialization, the forward pass, and result deserialization, representing the full processing cycle within the model server. The {Pod Response Latency} (0.02~s) captures the round-trip time from external request to response via KServe, under warmed-up conditions. The reported inference throughput (31800samp/s) is a theoretical maximum estimated under full compute utilization, assuming minimal external bottlenecks. Finally, the cold start time (62.02s) reflects the time from pod instantiation to readiness for serving the first batch, including model loading and pipeline initialization.

To better understand the responsiveness of our serving system, we provide a breakdown of inference and deployment-time metrics in Table~\ref{tab:timeMetrics}.  Different from the pipeline trace latency reported in Table~\ref{tab:runtime-breakdown}, which reflects the internal runtime of the inference module only, here we measured it at the KServe model server interface.  In a fully deployed environment, the end-to-end response time from an external request to receiving the prediction (via KServe) is 0.02 seconds, indicating that networking and orchestration overheads are minimal once the pod is active. The system can sustain an estimated inference throughput of 3180 samples per second, which offers an upper bound on what the system could handle under heavy workloads. Finally, we note that a cold start—encompassing container instantiation, model loading, and initialization —takes approximately 62 seconds.  Overall, these metrics jointly indicate that our pipeline supports responsive, continuous inference with modest compute overhead and low serving latency under realistic workloads.

%The MLP-based inference engine achieves 0.12s latency per sample, while pod-level end-to-end response time remains below 0.02s. With throughput exceeding 1300 samples per second and a cold start time about 32.02s, the system satisfies the runtime constraints required for continuous monitoring and just-in-time adaptation in production environments.

%% file: paper/Sec6-Conclusion.tex
%!TEX root = ../main.tex

\section{Conclusion and future directions}
\label{sec:conclusion}

As network softwarization continues to gain momentum, there is an urgent need for stable and predictable performance in the software data plane. Existing solutions for network performance diagnostics commonly rely on per-packet measurement for data collection, which requires tremendous engineering overhead and interferes with the critical data path. We propose a novel approach that utilizes low-level hardware features for KPI prediction. 
Compared to per-packet data collection, our approach is easily applicable to real-world NFV systems without an in-depth understanding of their implementation details. 
The low-level data collection imposes a negligible impact on the software data plane. We implement a tractable AI/ML model that can accurately infer throughput and latency in high-speed networks.
Our model is generalizable to network services with similar topological compositions, and its predictions can be interpreted with domain-specific knowledge to identify performance bottlenecks.

%\begin{table}[!t]
%\centering
%\scriptsize

%\begin{tabular}{lcccc}
%\toprule
%\textbf{Metric} & \textbf{Mean} & \textbf{Std Dev} & \textbf{Min} & \textbf{Max} \\
%\midrule
%JS Divergence ($js\_val$)     & 0.097 & 0.058 & 0.00 & 0.51 \\
%CPU Usage (\%)                & 21.97   & 10.19   & 7.20   & 62.70  \\
%Memory Usage (\%)             & 3.871   & 0.04    & 38.50  & 38.80  \\
%I/O Wait Time (\%)            & 0.25    & 0.22    & 0.00   & 1.00   \\
%Kafka Lag (msgs)              & 11.33   & 25.05   & 0.00   & 102.00 \\
%\bottomrule
%\end{tabular}
%\caption{Runtime metrics of the JS Divergence Monitor pod, collected via Prometheus. Statistics computed over one representative execution window (250 samples).}
%\label{tab:js-monitor-runtime}
%\end{table}
\begin{figure*}%[t]
  \centering
  \includegraphics[width=\linewidth]{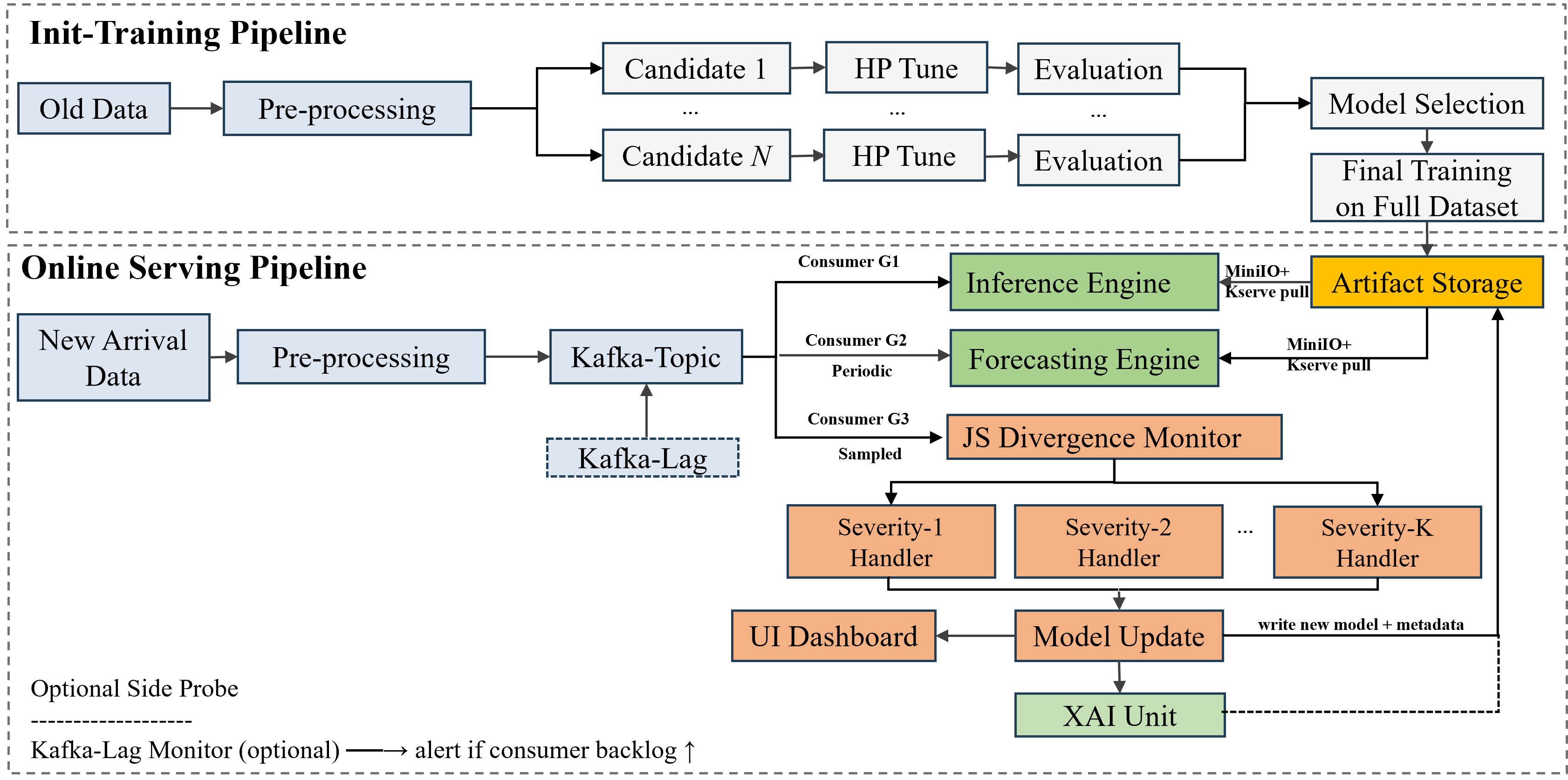}
  \caption{Infrastructure-level MLOps implementation DAG}
  \label{fig:app-platform}
\end{figure*}

%% file: paper/appendix.tex
\appendix
\section{End-to-end DAG implementation. }\label{app:infra}

The diagram records the concrete Kubeflow workflow instantiated in our cluster.   The upper branch depicts the \emph{init-training pipeline} and final artefact storage, while the lower branch corresponds to the \emph{online serving pipeline}, including inference, forecasting, drift detection, and adaptive update.  Background colours indicate functional layers:  
\colorbox{blue!18}{\strut blue} = data ingestion \& preprocessing,  
\colorbox{gray!25}{\strut gray} = offline model lifecycle,  
\colorbox{green!25}{\strut green} = long-running online services, and  
\colorbox{orange!22}{\strut orange} = monitoring \& self-adaptation.

\scriptsize{During the preparation of this work, the author(s) used OpenAI for minor language polishing. After using this tool, the authors reviewed and edited the content as needed and take full responsibility for the content of the publication.}